\documentclass[preprint,aps,prd,nofootinbib,amssymb,tightenlines,11pt]{revtex4-1}

\usepackage[russian,english]{babel}
\usepackage{amsthm}
\usepackage{exscale}
\usepackage{bbm,mathrsfs} 

\usepackage{bigdelim}

\providecommand{\bottomrule}{\botrule}
\providecommand{\midrule}{\colrule}
\usepackage[OT2,T1]{fontenc}
\usepackage{microtype}

\usepackage{graphicx}
\newcommand{\Graph}[2][1.0]{\vcenter{\hbox{\includegraphics[scale=#1]{Graphs/#2}}}}

\usepackage{mathtools}

\usepackage{numprint}

\usepackage[table]{xcolor}
\definecolor{links}{rgb}{0,0.5,0}

\usepackage[colorlinks=true,breaklinks=true,citecolor=blue!70!white]{hyperref}

\newcommand{\extrafootmark}[1]{\textsuperscript{\textcolor{red}#1}}

\DeclareMathAlphabet{\oldcal}{OMS}{cmsy}{m}{n}

\newcommand{\field}{\phi}
\newcommand{\n}{n}

\newcommand{\lagrangian}{\mathscr{L}}
\newcommand{\MS}{\ensuremath{\mathrm{MS}}}
\newcommand{\prim}{\ensuremath{\mathrm{prim}}}
\newcommand{\bigo}[1]{\oldcal{O}\left(#1\right)}
\newcommand{\D}{D}

\newcommand{\crit}{_{\star}}

\newcommand{\Rbar}{\R'}
\newcommand{\R}{\oldcal{R}}
\newcommand{\Rstar}{\R^{\ast}}

\newcommand{\Poles}{\K}
\newcommand{\K}{\oldcal{K}}

\newcommand{\mzv}[2][]{\zeta^{#1}_{#2}}

\newcommand{\defas}{\mathrel{\mathop:}=}

\newcommand{\setexp}[2]{\left\{ #1 \colon #2 \right\}}
\newcommand{\set}[1]{\left\{ #1 \right\}}
\newcommand{\abs}[1]{\left\vert#1\right\vert}
\newcommand{\restrict}[2]{\left.{#1}\right|_{#2}}

\newcommand{\Maple}{%
	\href{http://www.maplesoft.com/products/Maple/}
	{\textsf{\textup{Maple}}}%
}
\newcommand{\Mathematica}{%
	\href{http://www.wolfram.com/mathematica/}
	{\textsf{\textup{Mathematica}}}%
}
\newcommand{\MapleNote}{%
	\footnote{Maple is a trademark of Waterloo Maple Inc.}
}
\newcommand{\MapleTM}{%
	\href{http://www.maplesoft.com/products/Maple/}
	{\textsf{\textup{Maple}}\texttrademark}%
}
\newcommand{\Python}{%
	\href{https://www.python.org/}
	{\textsf{\textup{Python}}}%
}

\newcommand{\NI}[1]{\mathrm{NI}\left( #1 \right)}
\newcommand{\Period}[1]{\mathcal{P}\left( #1 \right)}
\newcommand{\psipol}{\psi}
\newcommand{\SP}{x}
\newcommand{\Hepp}[1]{\mathcal{H}\left( #1 \right)}
\DeclareMathOperator{\Aut}{Aut}
\newcommand{\GroupFactor}[1]{\mathscr{C}_{#1}}
\newcommand{\Sym}[1]{\mathrm{Sym}\left( #1 \right)}

\newcommand{\dd}[1][]{\mathrm{d}^{#1}}
\newcommand{\eps}{\varepsilon}

\newcommand{\Q}{\mathbbm{Q}}

\newcommand{\EM}{\gamma_{\text{\upshape{E}}}}
\newcommand{\GK}{A}

\newcommand{\Variation}[1]{\mathrm{Var}_{#1}}

\newcommand{\Borel}[2][]{\mathfrak{B}^{#1}_{#2}}

\newcommand{\File}[1]{\texttt{\textup{#1}}}
\newcommand{\JaxoDraw}{\texttt{\textup{JaxoDraw}}}
\newcommand{\Axodraw}{\texttt{\textup{Axodraw}}}
\newcommand{\HyperInt}{\href{http://bitbucket.org/PanzerErik/hyperint/}{\texttt{\textup{HyperInt}}}}
\newcommand{\feyngen}{\href{https://github.com/michibo/feyncop}{\texttt{\textup{feyngen}}}}

\newcommand{\GraphState}{\texttt{\textup{GraphState}}}
\newcommand{\Graphine}{\texttt{\textup{Graphine}}}
\newcommand{\nauty}{\href{http://pallini.di.uniroma1.it/}{\texttt{\textup{nauty}}}}
\newcommand{\FORM}{\href{https://www.nikhef.nl/~form/}{\textsc{Form}}}

\theoremstyle{plain}
	\newtheorem{theorem}{Theorem}
\theoremstyle{definition}
	\newtheorem{definition}[theorem]{Definition}
\theoremstyle{remark}

\begin{document}
\title{Minimally subtracted six loop renormalization of $O(\n)$-symmetric $\field^4$ theory and critical exponents}

\author{Mikhail V.~Kompaniets}
\email{m.kompaniets@spbu.ru}
\affiliation{St.~Petersburg State University, 7/9 Universitetskaya nab., St.~Petersburg 199034, Russia}

\author{Erik Panzer}
\email{erik.panzer@all-souls.ox.ac.uk}
\affiliation{All Souls College, University of Oxford, OX1 4AL Oxford, UK}

\begin{abstract}
	We present the perturbative renormalization group functions of $O(\n)$-symmetric $\field^4$ theory in $4-2\eps$ dimensions to the sixth loop order in the minimal subtraction scheme.
	In addition, we estimate diagrams without subdivergences up to 11 loops and compare these results with the asymptotic behaviour of the beta function.
	Furthermore we perform a resummation to obtain estimates for critical exponents in three and two dimensions.
\end{abstract}

\maketitle

\section{Introduction}
\label{sec:intro}%

The field-theoretic renormalization group approach \cite{Vasilev,ZinnJustin,KleinertSchulteFrohlinde:CriticalPhi4} has a long and successful history in the study of critical phenomena, going back to the famous $\eps$-expansion \cite{WilsonFisher:3.99}.
In particular, it predicts critical exponents of second order phase transitions with high accuracy \cite{LeGuillouZinnJustin:n3D} when combined with resummation methods \cite{LeGuillouZinnJustin:CriticalFromField}.
More specifically, one can extract approximate exponents for three dimensional $O(\n)$ universality classes from the renormalization group functions of $\field^4$ theory in $4-2\eps$ dimensions.\footnote{%
	Another approach renormalizes the theory directly in three dimensions \cite{LeGuillouZinnJustin:n3D,BakerNickelGreenMeiron:Ising3}.
}
Considerable effort has thus been invested in the calculation of the latter to increasingly high orders in perturbation theory.

After the results \cite{BrezinLeGuillouZinnJustinNickel:HigherOrder,VladimirovKazakovTarasov:Calculation} for three and four loops, the computation \cite{ChetyrkinKataevTkachov:5loopPhi4,ChetyrkinGorishnyLarinTkachov:5loopPhi4,Kazakov:MethodOfUniqueness} of the fifth order yielded highly accurate critical exponents \cite{GuillouZinnJustin:Accurate}.
The subsequent correction \cite{KNFCL:5loopPhi4} of the perturbative result, which only very recently was confirmed by numeric methods \cite{AdzhemyanKompaniets:5loopNumerical}, affected the resummed exponents only marginally \cite{GuidaZinnJustin:CriticalON}.

For many years, the renormalization group method in $3$ dimensions provided the most accurate theoretical predictions for critical exponents, consistent with the only slightly less precise results from the fifth order $\eps$-expansion \cite{GuidaZinnJustin:CriticalON}.
However, considerable progress of other techniques has by now produced a multitude of much more refined results. Among those, we like to point out the particularly astonishing performance of the conformal bootstrap program \cite{EcheverriHarlingSerone:EffectiveBootstrap,KosPolandDuffinVichi:Islands} and Monte Carlo methods \cite{ClisbyDuenweg:Hydrodynamic,Clisby:ScaleFreeSAW}, which reached unprecedented accuracy in some cases.

It was therefore overdue to improve on the $\eps$-expansion, which had been stuck at five loops for 25 years.
Finally, the 6-loop result for the field anomalous dimension was published in \cite{BatkovichKompanietsChetyrkin:6loop}, and we provided the complete set of renormalization group functions for $\field^4$ theory ($\n=1$) in \cite{KompanietsPanzer:LL2016}. 
These were obtained in a Feynman diagram computation, that became feasible through the automatization of new techniques \cite{Panzer:HyperIntAlgorithms,Brown:TwoPoint,BrownKreimer:AnglesScales} to calculate Feynman integrals, very briefly summarized in section~\ref{sec:computation}.

Here, we present the six loop renormalization group functions for arbitrary values of $\n$, in the minimal subtraction scheme. The exact (and slightly unwieldy) expressions are given in section~\ref{sec:RG-functions}, together with tables of numeric values for the most interesting cases $\n\in\set{0,1,2,3,4}$.
We then discuss numerous checks of our result and like to stress in particular the confirmation of the beta function and the field anomalous dimension, at $\n=1$, by Oliver Schnetz \cite{Schnetz:NumbersAndFunctions}.
Furthermore, we compare the coefficients of the beta function, supplemented by estimates up to 11 loops, with the expected asymptotic behaviour.
This analysis confirms the known fact that the convergence is rather slow in absolute terms, but it also shows that the qualitative trend is correct and we observe a striking pattern of zeros in the dependence on $\n$, in agreement with the asymptotic prediction.

In section~\ref{sec:resummation}, we give the $\eps$-expansions for the critical exponents and recall the Borel resummation method with conformal mapping, which was employed to great effect in \cite{VladimirovKazakovTarasov:Calculation,GuillouZinnJustin:Accurate}. Since this basic idea can be implemented in many different ways and incorporates several arbitrary parameters, we include a rather detailed discussion of its characteristics.
Our resummation algorithm is accurately defined in section~\ref{sec:resummation-algorithm}, followed by a discussion of how we estimate the errors.

Finally, our resummed results for the critical exponents in three and two dimensions are summarized and discussed in sections~\ref{sec:3dim-results} and \ref{sec:2dim-results}. The reader only interested in these results will find them in tables~\ref{tab:exponents-3d} and \ref{tab:exponents-2d}. In short, we find increased accuracy (in comparison to the five loop resummation) and good agreement with results from other methods.
While the record precision for $\eta$ and $\nu$ in the cases $\n=0$ and $\n=1$ from bootstrap and Monte Carlo methods is clearly out of reach, the $\eps$-expansion seems superior for the correction to scaling exponent $\omega$ and is on par for the Fisher exponent $\eta$ when $\n \geq 2$. Values for $\nu$ tend to be low in comparison with simulations in $3$ dimensions and the theoretical predictions in $2$ dimensions.

In our conclusion~\ref{sec:outlook}, we anticipate that the upcoming $7$-loop renormalization \cite{Schnetz:NumbersAndFunctions} is very likely to result in estimates with smaller uncertainties, which will provide even stronger tests of the compatibility of different theoretical approaches.

We hope that our results will be useful for further analyses. In particular, it would be interesting to compare our critical exponents with other resummation methods applied to the six loop series.
Another application might be to probe the asymptotic behaviour of the renormalization group functions, as in \cite{Shrock:6loopPhi4,Shrock:BetaZeroPhi4}.
Also, the $\eps$-expansions and $Z$-factor contributions of individual diagrams should suffice to study other universality classes like the $O(\n)$ model with cubic anisotropy \cite{MudrovVarnashev:ModifiedBorel,KleinertSchulteFrohlinde:CubicEps5} or even more complicated cases like \cite{CalabreseParrucciniSolokov:Chiral,KalagovKompanietsNalimov:U(r)}.

Therefore we provide an extensive set of data with this article as described in \appendixname~\ref{sec:files} and available under \href{https://doi.org/10.5287/bodleian:pvx4nMyQr}{DOI 10.5287/bodleian:pvx4nMyQr}. We added the computed Feynman integrals to the \href{http://www.loopedia.net/}{\sffamily{Loopedia}} \cite{Loopedia:Release}.

\section{Field theory and renormalization}
\label{sec:phi4theory}

We consider the theory of $\n$ scalar fields $\field=(\field_1,\ldots,\field_{\n})$ with an $O(\n)$ symmetric interaction $\field^4 \defas (\phi^2)^2 = (\field_1^2+\cdots+\field_{\n}^2)^2$.
In $\D=4-2\eps$ Euclidean dimensions, the corresponding renormalized Lagrangian is
\begin{equation}
	\lagrangian
	=
		\frac{1}{2}m^2 Z_1\field^2 
		+\frac{1}{2}Z_2\left(\partial\field \right)^2
		+\frac{16\, \pi^2}{4!}Z_4 \, g\, \mu^{2\eps}\, \field^4
	\label{eq:lagrangian}%
\end{equation}
and contains an arbitrary mass scale $\mu$, such that $g$ stays dimensionless. The $Z$-factors relate the renormalized field $\field$, mass $m$ and coupling $g$ to the bare field $\field_0$, bare mass $m_0$ and bare coupling $g_0$ via
\begin{equation}\label{eq:Z-factors}%
	Z_{\field}
		= \frac{\field_0}{\field} 
		= \sqrt{Z_2}
	,\quad
	Z_{m^2} 
		= \frac{m_0^2}{m^2}
		= \frac{Z_1}{Z_2}
	\quad\text{and}\quad
	Z_g
		= \frac{g_0}{\mu^{2\eps} g}
		= \frac{Z_4}{Z_2^2}
	.
\end{equation}
In dimensional regularization \cite{tHooftVeltman:RegularizationGaugeFields} and minimal subtraction, these $Z$-factors depend only on $\eps$ and $g$ and admit expansions into formal Laurent series
\begin{equation}
	Z_i = Z_i(g,\eps)
	= 1 + \sum_{k=1}^{\infty} \frac{Z_{i,k}(g)}{\eps^k}
	,
	\label{eq:Z-factor-expansion}%
\end{equation}
where each $Z_{i,k}(g)$ is a formal power series in the coupling $g$.
The renormalization group (RG) functions can be read off from the residues (at $\eps=0$) of the $Z$-factors \cite{tHooft:DimRegRG,Collins:CountertermsDimReg}. In particular, the beta function can be computed as
\begin{equation}
	\beta(g,\eps)
	\defas \restrict{
		\mu\frac{\partial g}{\partial \mu}
	}{g_0}
	= - 2\eps \left( \frac{\partial \log (g Z_g)}{\partial g} \right)^{-1}
	= -2\eps g + 2 g^2 \frac{\partial Z_{g,1}(g)}{\partial g}
	,
	\label{eq:beta-def}%
\end{equation}
whereas the anomalous dimensions for the field and mass are given by
\begin{equation}
	\gamma_{i}(g) 
	\defas \restrict{
		\mu \frac{\partial \log Z_{i}}{\partial \mu}
	}{g_0,m_0,\field_0}
	= \beta(g) \frac{\partial \log Z_i(g)}{\partial g}
	= -2 g \frac{\partial Z_{i,1}(g)}{\partial g}
	\quad\text{for}\quad
	i=m^2, \field
	.
	\label{eq:anom-dims}%
\end{equation}
These RG functions are formal power series in the coupling $g$. The first terms
\begin{equation*}
	\beta(g,\eps)
	= -2\eps g + \frac{\n+8}{3} g^2 + \bigo{g^3}
\end{equation*}
show that, at least for small $\eps>0$, the beta function admits a non-trivial zero 
\begin{equation}
	g\crit(\eps)
	= \frac{6}{\n+8} \eps + \bigo{\eps^2}
	\quad\text{such that}\quad
	\beta(g\crit(\eps),\eps)=0.
	\label{eq:gcrit-first-order}%
\end{equation}
This \emph{critical coupling} is a formal power series in $\eps$ and determines a fixed point of the renormalization group flow. This fixed point is IR-attractive, meaning that the \emph{correction to scaling exponent}
\begin{equation}
	\omega(\eps)
	\defas \beta'(g\crit(\eps),\eps)
	= \restrict{\frac{\partial}{\partial g}}{g=g\crit(\eps)} \beta(g,\eps)
	= 2\eps + \bigo{\eps^2}
	\label{eq:def-omega}%
\end{equation}
is positive.
The anomalous dimensions at the critical point define the \emph{critical exponents}
\begin{equation}
	\eta(\eps)
	\defas 2 \gamma_{\field} (g\crit(\eps))
	\quad\text{and}\quad
	\nu(\eps)
	\defas \left[ 2 + \gamma_{m^2} (g\crit(\eps)) \right]^{-1}
	,
	\label{eq:def-eta-nu}%
\end{equation}
which we compute as formal power series in $\eps$. According to the leading terms
\begin{equation}
	\gamma_{\field}
	= \frac{\n+2}{36} g^2 + \bigo{g^3}
	\quad\text{and}\quad
	\gamma_{m^2}
	= - \frac{\n+2}{3} g + \bigo{g^2}
	,
	\label{eq:anom-first-order}%
\end{equation}
the first terms of their $\eps$-expansions are
\begin{equation*}
	\eta(\eps)
	= \frac{2(\n+2)\eps^2}{(\n+8)^2} + \bigo{\eps^3}
	\quad\text{and}\quad
	\nu(\eps)
	= \frac{1}{2} + \frac{(\n+2)\eps}{2(\n+8)} + \bigo{\eps^2}
	.
\end{equation*}
We note that there are more critical exponents, but those are related to $\eta$ and $\nu$ via the following \emph{scaling{\footnotemark} relations} \cite{KleinertSchulteFrohlinde:CriticalPhi4}, which, for the purpose of this paper, we simply take as definitions of $\alpha$, $\beta$, $\gamma$ and $\delta$:
\footnotetext{%
	Relations that explicitly involve the dimension $\D$ are often distinguished and called \emph{hyperscaling relations}. For simplicity, we will refer to all of \eqref{eq:scaling-relations} just as \emph{scaling relations}.
}
\begin{equation}
	\gamma=\nu(2-\eta)
	,\quad
	\D \nu=2-\alpha
	,\quad
	\beta\delta=\beta+\gamma
	\quad\text{and}\quad
	\alpha+2\beta+\gamma=2
	.
	\label{eq:scaling-relations}%
\end{equation}
The critical exponents and the correction to scaling exponent $\omega$ are independent of the renormalization scheme and they conjecturally describe phase transitions of numerous physical systems in several universality classes.
In section~\ref{sec:resummation} we describe how we resummed the $\eps$-expansions to arrive at the estimates for these quantities in $\D=3$ and $\D=2$ dimensions as presented in sections~\ref{sec:3dim-results} and \ref{sec:2dim-results}.

\section{Calculational techniques}
\label{sec:computation}

We compute the $Z$-factors \eqref{eq:Z-factors} as the counterterms for the one-particle irreducible correlation functions $\Gamma_N$ of $N=2$ and $N=4$ fields, by expanding them as Feynman diagrams.
The ultraviolet (UV) subdivergences are subtracted with the Bogoliubov-Parasiuk $\Rbar$-operation \cite{BogoliubovParasiuk:Kausalfunktionen,BogShirk}, such that
\begin{equation}\label{eq:Z-from-R}
\begin{split}
	Z_1 &= 1 + \partial_{m^2} \Poles \Rbar {\Gamma}_2 (p,m^2,g,\mu)
	, \\ 
	Z_2 &= 1 + \partial_{p^2} \Poles \Rbar {\Gamma}_2 (p,m^2,g,\mu)
	\quad\text{and} \\ 
	Z_4 &= 1 + \Poles \Rbar {\Gamma}_4 (p,m^2,g,\mu)/g
	.
\end{split}
\end{equation}
In the minimal subtraction (\MS) scheme, the form \eqref{eq:Z-factor-expansion} of the $Z$-factors is obtained by projecting onto the pole part with respect to the regulator $\eps=(4-\D)/2$,
\begin{equation}
	\Poles \left( \sum_n c_n \eps^n \right)
	\defas
	\sum_{n<0} c_n \eps^n
	.
	\label{eq:Poles-def}%
\end{equation}
We use standard techniques \cite{Vasilev,ZinnJustin,KleinertSchulteFrohlinde:CriticalPhi4} to simplify the computation:
\begin{itemize}
	\item Acting with $-\partial_{m^2}$ squares a propagator, which is equivalent to a $4$-point graph with two vanishing external momenta. This way, $Z_{1}$ can be expressed in terms of a subset of the graphs contributing to $Z_4$ \cite[section~11.7]{KleinertSchulteFrohlinde:CriticalPhi4}.

	\item Using infrared rearrangement (IRR) \cite{Vladimirov:ManyLoopPhi4,ChetyrkinKataevTkachov:Gegenbauer}, we can set all internal masses to zero and nullify some external momenta such that only massless propagators remain to be computed. 
\end{itemize}
These express all $Z$-factors in terms of $p$-integrals (massless propagators) without infrared divergences, and our task is thus reduced to the computation of the $\eps$-expansion of these integrals.
The number of $\field^4$ Feynman graphs contributing to $\Gamma_2$ and $\Gamma_4$ is summarized in table~\ref{tab:graph-counts}.\footnote{%
	The $\field^4$ graphs with $\leq 6$ loops were already enumerated and tabulated in \cite{NickelMeironBaker:Compilation24}. 
	Asymptotically, the number of graphs grows factorially with the number of loops and precise higher order expansions were recently presented in \cite{Borinsky:RenormalizedAsymptoticEnumeration}.
}
A few of them are primitive (free of subdivergences) and those were computed, up to $\leq 6$ loops, already long ago in \cite{Broadhurst:5loopsbeyond}.\footnote{%
	Some of the results in \cite{Broadhurst:5loopsbeyond} in terms of zeta values were based on numeric techniques, and one value in particular was only later identified in \cite{BroadhurstKreimer:KnotsNumbers} as the double zeta value \eqref{eq:zeta35}.
	Analytic proofs of these $6$-loop periods were later given in \cite{Schnetz:K34,Panzer:MasslessPropagators,Schnetz:GraphicalFunctions}.
}
In fact, the partial results \cite{BroadhurstKreimer:KnotsNumbers,Schnetz:Census} at higher loop orders have recently been augmented significantly, including in particular the complete set of primitive $\field^4$ Feynman integrals with $7$ loops \cite{PanzerSchnetz:Phi4Coaction}.

In order to calculate the missing integrals with (UV) subdivergences, which was the main technical challenge, we construct auxiliary counterterms using the $\Rbar$ operation of the BPHZ-like \emph{one-scale scheme} introduced in \cite{BrownKreimer:AnglesScales}. The resulting linear combinations of integrals are convergent in $\D=4$ dimensions and can thus be computed exactly, term-by-term after expanding in $\eps$, with the program {\HyperInt} \cite{Panzer:HyperIntAlgorithms} based on the algorithm proposed in \cite{Brown:TwoPoint}.

We gave a detailed account of this new method in \cite{KompanietsPanzer:LL2016}. The entire computation is automated with programs written in {\MapleTM} and {\Python}, using the {\GraphState}/{\Graphine} library to manipulate Feynman graphs \cite{BatkovichKirienkoKompanietsNovikov:GraphState,BatkovichKompaniets:Toolbox}, which will be published separately.\MapleNote

The only addition to be made to the exposition in \cite{KompanietsPanzer:LL2016}, is that each Feynman graph $G$ is now not only weighted with the usual combinatorial symmetry factor $1/\abs{\Aut(G)}$, but also with an additional $O(\n)$-\emph{group factor} $\GroupFactor{G}$ which is a polynomial in $\n$.
It equals the number of ways one can assign a component $1 \leq i_e \leq \n$ of the field $\field$ to each of the edges $e$ of the graph, in such a way that at each vertex $v$, the flavours of the four edges $v^{[1]},\ldots,v^{[4]}$ meeting at $v$ can be grouped into two equal pairs---according to the interaction term $(\field^2)^2$. Hence,
\begin{equation}
	\GroupFactor{G}
	= \sum_{i_1,\ldots,i_{E(G)}=1}^{n} \prod_{v\in V(G)} \lambda\left(i_{v^{[1]}},\ldots i_{v^{[4]}}\right)
	\label{eq:group-factor}%
\end{equation}
gives the group factor for vacuum (i.e.\ 4-regular) graphs, with
\begin{equation}
	\lambda(a,b,c,d) 
	\defas \frac{\delta_{a,b}\delta_{c,d}+\delta_{a,c}\delta_{b,d}+\delta_{a,d}\delta_{b,c}}{3}
	.
	\label{eq:vertex-delta}%
\end{equation}
For a propagator (2-point) graph $G$, let $\widehat{G}$ denote the vacuum graph obtained by gluing the external legs together, and for a vertex (4-point) graph $G$, let $\widehat{G}$ denote the graph obtained by attaching all external legs to an additional vertex (this is known as the \emph{completed} graph \cite{Schnetz:Census}). Then it is easy to check that
\begin{equation}
	\GroupFactor{G}
	= \GroupFactor{\widehat{G}} \cdot
	\begin{cases}
		\frac{1}{\n} & \text{if $G$ is a propagator graph, and} \\
		\frac{3}{\n(\n+2)} & \text{if $G$ is a vertex graph.} \\
	\end{cases}
	\label{eq:group-factor-from-vacuum}%
\end{equation}

\begin{table}
	\centering%
	\caption{The number of (isomorphism classes of) one-particle irreducible $\field^4$ Feynman graphs, excluding graphs with tadpoles (as those vanish in massless DimReg). We include the column for $7$ loops just to illustrate the growth in complexity.}%
	\label{tab:graph-counts}%
	\begin{tabular}{rrrrrrrr}
	\toprule
		loops                                 & 1 & 2 & 3 &  4 &   5 &   6 &    7 \\
	\midrule
		2-point graphs ($\Gamma_2$)           & 0 & 1 & 1 &  4 &  11 &  50 &  209\\ 
		4-point graphs ($\Gamma_4$)           & 1 & 2 & 8 & 26 & 124 & 627 & 3794 \\
		primitive 4-point graphs ($\Gamma_4$) & 1 & 0 & 1 &  1 &   3 &  10 &   44\\
	\bottomrule
	\end{tabular}%
\end{table}

\section{Results for the RG functions}
\label{sec:RG-functions}

We now present our results for the $6$-loop renormalization group functions, computed in the minimal subtraction (MS) scheme in $\D=4-2\eps$ dimensions.
Among those, the anomalous dimension of the field takes the simplest form, because it only involves Riemann zeta values $\mzv{k} = \sum_{n=1}^{\infty} 1/n^k$:
\begin{equation}\begin{split}
\label{eq:gamma-phi}%
\gamma^{\MS}_{\phi}(g) &=
\frac{n+2}{36} g^2
-\frac{(n+8)(n+2)}{432} g^3
-\frac{5(n^2-18n-100)(n+2)}{\numprint{5184}}g^4
\\&
-\Big[
	\numprint{1152}
	(
		5\n
		+22
	) \mzv{4}
	-\numprint{48}
	(
		\n^{3}
		-6\n^{2}
		+64\n
		+184
	) \mzv{3}
\\&\quad
	+
	(
		39\n^{3}
		+296\n^{2}
		+\numprint{22752}\n
		+\numprint{77056}
	) 
\Big] \frac{(n+2) g^{5}}{\numprint{186624}}
\\&
-\Big[
	\numprint{512}
	(
		2\n^{2}
		+55\n
		+186
	) \mzv[2]{3}
	-\numprint{6400}
	(
		2\n^{2}
		+55\n
		+186
	) \mzv{6}
\\&\quad
	+\numprint{4736}
	(
		\n
		+8
	)
	(
		5\n
		+22
	) \mzv{5}
	-\numprint{48}
	(
		\n^{4}
		+2\n^{3}
		+328\n^{2}
		+\numprint{4496}\n
		+\numprint{12912}
	) \mzv{4}
\\&\quad
	+\numprint{16}
	(
		\n^{4}
		-936\n^{2}
		-\numprint{4368}\n
		-\numprint{18592}
	) \mzv{3}
\\&\quad
	+
	(
		29\n^{4}
		+794\n^{3}
		-\numprint{30184}\n^{2}
		-\numprint{549104}\n
		-\numprint{1410544}
	) 
\Big] \frac{(n+2)g^{6}}{\numprint{746496}}
+\bigo{g^{7}}
\end{split}
\end{equation}
\begin{table}
	\centering%
	\caption{Numerical values for the $6$-loop field anomalous dimension.}%
	\label{tab:gamma_phi-numeric}%
	\begin{tabular}{cc}
	\toprule
		$\quad\n\quad$ & $\gamma^{\MS}_{\field}(g)$   \\
	\midrule
		$0$ & $0.05556g^{2} -0.03704g^{3} +0.1929g^{4} -1.006g^{5} +7.095g^{6}+\bigo{g^{7}}$ \\
		$1$ & $0.08333g^{2} -0.06250g^{3} +0.3385g^{4} -1.926g^{5} +14.38g^{6}+\bigo{g^{7}}$ \\
		$2$ & $0.11111g^{2} -0.09259g^{3} +0.5093g^{4} -3.148g^{5} +24.71g^{6}+\bigo{g^{7}}$ \\
		$3$ & $0.13889g^{2} -0.12731g^{3} +0.6993g^{4} -4.689g^{5} +38.44g^{6}+\bigo{g^{7}}$ \\
		$4$ & $0.16667g^{2} -0.16667g^{3} +0.9028g^{4} -6.563g^{5} +55.93g^{6}+\bigo{g^{7}}$ \\
	\bottomrule
	\end{tabular}%
\end{table}%
Note that this result was already obtained in \cite{BatkovichKompanietsChetyrkin:6loop} with different methods; so our computation provides a non-trivial check. Numeric values for $\n\in\set{0,1,2,3,4}$ are given in table~\ref{tab:gamma_phi-numeric}.
The following results contain also the double zeta value
\begin{equation}
	\mzv{3,5}
	\defas
	\sum_{1\leq n < m} \frac{1}{n^3 m^5}
	\approx 
	0.037707673
	,
	\label{eq:zeta35}%
\end{equation}
which (conjecturally) cannot be written as a polynomial in Riemann zeta values with rational coefficients \cite{Broadhurst:1440,Broadhurst:5loopsbeyond}. 
Our new results are the coefficient of $g^6$ in the anomalous dimension of the mass (see table~\ref{tab:gamma_m^2-numeric} for numeric expansions),
\begin{equation}\begin{split}
\label{eq:gamma-m^2}%
\gamma^{\MS}_{m^2}(g) &=
-\frac{(n+2)g}{3}
+\frac{5(n+2) g^{2}}{18}
- \frac{(5n+37)(n+2)g^3}{36}
\\&
+\Big[
	\numprint{288}
	(
		5\n
		+22
	) \mzv{4}
	+\numprint{48}
	(
		3\n^{2}
		+10\n
		+68
	) \mzv{3}
	-
	(
		\n^{2}
		-\numprint{7578}\n
		-\numprint{31060}
	) 
\Big] \frac{(n+2) g^{4}}{\numprint{7776}}
\\&
-\Big[
	\numprint{9600}
	(
		2\n^{2}
		+55\n
		+186
	) \mzv{6}
	-\numprint{768}
	(
		2\n^{2}
		+145\n
		+582
	) \mzv[2]{3}
	-\numprint{768}
	(
		5\n^{2}
		-14\n
		-72
	) \mzv{5}
\\&\quad
	-\numprint{288}
	(
		3\n^{3}
		-29\n^{2}
		-816\n
		-\numprint{2668}
	) \mzv{4}
	+\numprint{48}
	(
		17\n^{3}
		+940\n^{2}
		+\numprint{8208}\n
		+\numprint{31848}
	) \mzv{3}
\\&\quad
	+
	(
		21\n^{3}
		+\numprint{45254}\n^{2}
		+\numprint{1077120}\n
		+\numprint{3166528}
	) 
\Big] \frac{(n+2) g^{5}}{\numprint{186624}}
\\&
+\Big[
	\numprint{1152}
	(
		14\n^{2}
		+189\n
		+526
	)
	(
		2063\mzv{8}
		-144\mzv{3,5}
	)
	-\numprint{921600}
	(
		15\n^{2}
		+239\n
		+718
	) \mzv{3}\mzv{5}
\\&\quad
	-\numprint{640}
	(
		\numprint{1080}\n^{3}
		+\numprint{25180}\n^{2}
		+\numprint{284525}\n
		+\numprint{814062}
	) \mzv{7}
\\&\quad
	-\numprint{15360}
	(
		2\n^{3}
		-157\n^{2}
		-\numprint{2512}\n
		-\numprint{8268}
	) \mzv{3}\mzv{4}
\\&\quad
	+\numprint{5120}
	(
		27\n^{3}
		+\numprint{1082}\n^{2}
		+\numprint{13072}\n
		+\numprint{40008}
	) \mzv[2]{3}
\\&\quad
	+\numprint{3200}
	(
		285\n^{3}
		+\numprint{7178}\n^{2}
		+\numprint{73768}\n
		+\numprint{196032}
	) \mzv{6}
\\&\quad
	+\numprint{320}
	(
		45\n^{4}
		+\numprint{3622}\n^{3}
		+\numprint{12202}\n^{2}
		+\numprint{207708}\n
		+\numprint{753040}
	) \mzv{5}
\\&\quad
	-\numprint{240}
	(
		47\n^{4}
		+\numprint{2606}\n^{3}
		-\numprint{5480}\n^{2}
		-\numprint{194320}\n
		-\numprint{489328}
	) \mzv{4}
\\&\quad
	-\numprint{80}
	(
		51\n^{4}
		-\numprint{9208}\n^{3}
		-\numprint{419076}\n^{2}
		-\numprint{3342688}\n
		-\numprint{8997136}
	) \mzv{3}
\\&\quad
	-\numprint{5}
	(
		43\n^{4}
		+\numprint{48234}\n^{3}
		-\numprint{5154216}\n^{2}
		-\numprint{63140784}\n
		-\numprint{145482928}
	) 
\Big] \frac{(n+2) g^{6}}{\numprint{9331200}}
\\&
+\bigo{g^{7}}
\end{split}
\end{equation}
\begin{table}
	\centering%
	\caption{Numerical values for the $6$-loop mass anomalous dimension.}%
	\label{tab:gamma_m^2-numeric}%
	\begin{tabular}{cc}
	\toprule
		$\quad\n\quad$ & $\gamma_{m^2}^{\MS}(g)$   \\
	\midrule
		$0$ & $ -0.6667g +0.5556g^{2} -2.056g^{3} +10.76g^{4} -75.70g^{5} +636.7g^{6}+\bigo{g^{7}}$ \\
		$1$ & $ -1.0000g +0.8333g^{2} -3.500g^{3} +19.96g^{4} -150.8g^{5} +1355g^{6}+\bigo{g^{7}}$ \\
		$2$ & $ -1.3333g +1.1111g^{2} -5.222g^{3} +31.87g^{4} -255.8g^{5} +2434g^{6}+\bigo{g^{7}}$ \\
		$3$ & $ -1.6667g +1.3889g^{2} -7.222g^{3} +46.64g^{4} -394.9g^{5} +3950g^{6}+\bigo{g^{7}}$ \\
		$4$ & $ -2.0000g +1.6667g^{2} -9.500g^{3} +64.39g^{4} -571.9g^{5} +5983g^{6}+\bigo{g^{7}}$ \\
	\bottomrule
	\end{tabular}%
\end{table}%
and the coefficient of $g^7$ in the beta function
\begin{equation}\begin{split}
\label{eq:betaN}%
\beta^{\MS}(g) &=
- 2 \eps g
+\frac{\n+8}{3}g^2
-\frac{3 \n +14}{3}g^3
+\frac{
    96 (5 \n+22) \mzv{3}
  + 33 \n^2 +922 \n + \numprint{2960}
}{216}g^4 
\\&
-\Big[
	\numprint{1920} (
		2 \n^2
		+55 \n
		+186 
	) \mzv{5}
	-288 (
		\n
		+8 
	) (
		5 \n
		+22 
	) \mzv{4}
\\&\quad
	+96 (
		63 \n^2
		+764 \n
		+\numprint{2332} 
	) \mzv{3}
	- (
		5 \n^3
		-\numprint{6320} \n^2
		-\numprint{80456} \n
		-\numprint{196648} 
	) 
\Big] \frac{g^5}{\numprint{3888}}
\\&
+\Big[
	\numprint{112896} (
		14 \n^2
		+189 \n
		+526 
	) \mzv{7}
	-768 (
		6 \n^3
		+59 \n^2
		-446 \n
		-\numprint{3264} 
	) \mzv[2]{3}
\\&\quad
	-\numprint{9600} (
		\n
		+8 
	) (
		2 \n^2
		+55 \n
		+186 
	) \mzv{6}
	+256 (
		305 \n^3
		+\numprint{7466} \n^2
		+\numprint{66986} \n
		+\numprint{165084} 
	) \mzv{5}
\\&\quad
	-288 (
		63 \n^3
		+\numprint{1388} \n^2
		+\numprint{9532} \n
		+\numprint{21120} 
	) \mzv{4}
\\&\quad
	-16 (
		9 \n^4
		-\numprint{1248} \n^3
		-\numprint{67640} \n^2
		-\numprint{552280} \n
		-\numprint{1314336} 
	) \mzv{3}
\\&\quad
	+ (
		13 \n^4
		+\numprint{12578} \n^3
		+\numprint{808496} \n^2
		+\numprint{6646336} \n
		+\numprint{13177344} 
	) 
\Big] \frac{g^6}{\numprint{62208}}
\\&
-\Big[
	\numprint{204800} (
		\numprint{1819} \n^3
		+\numprint{97823} \n^2
		+\numprint{901051} \n
		+\numprint{2150774} 
	) \mzv{9}
\\&\quad
	+\numprint{14745600} (
		\n^3
		+65 \n^2
		+619 \n
		+\numprint{1502} 
	) \mzv[3]{3}
\\&\quad
	+\numprint{995328} (
		42 \n^3
		+\numprint{2623} \n^2
		+\numprint{25074} \n
		+\numprint{59984} 
	) \mzv{3,5}
\\&\quad
	-\numprint{20736} (
		\numprint{28882} \n^3
		+\numprint{820483} \n^2
		+\numprint{6403754} \n
		+\numprint{14174864} 
	) \mzv{8}
\\&\quad
	-\numprint{5529600} (
		8 \n^3
		-635 \n^2
		-\numprint{9150} \n
		-\numprint{25944} 
	) \mzv{3}\mzv{5}
\\&\quad
	+\numprint{11520} (
		440 \n^4
		+\numprint{126695} \n^3
		+\numprint{2181660} \n^2
		+\numprint{14313152} \n
		+\numprint{29762136} 
	) \mzv{7}
\\&\quad
	+\numprint{207360} (
		\n
		+8 
	) (
		6 \n^3
		+59 \n^2
		-446 \n
		-\numprint{3264} 
	) \mzv{3}\mzv{4}
\\&\quad
	-\numprint{23040} (
		188 \n^4
		+132 \n^3
		-\numprint{93363} \n^2
		-\numprint{862604} \n
		-\numprint{2207484} 
	) \mzv[2]{3}
\\&\quad
	-\numprint{28800} (
		595 \n^4
		+\numprint{20286} \n^3
		+\numprint{277914} \n^2
		+\numprint{1580792} \n
		+\numprint{2998152} 
	) \mzv{6}
\\&\quad
	+\numprint{5760} (
		\numprint{4698} \n^4
		+\numprint{131827} \n^3
		+\numprint{2250906} \n^2
		+\numprint{14657556} \n
		+\numprint{29409080} 
	) \mzv{5}
\\&\quad
	+\numprint{2160} (
		9 \n^5
		-\numprint{1176} \n^4
		-\numprint{88964} \n^3
		-\numprint{1283840} \n^2
		-\numprint{6794096} \n
		-\numprint{12473568} 
	) \mzv{4}
\\&\quad
	-720 (
		33 \n^5
		+\numprint{2970} \n^4
		-\numprint{477740} \n^3
		-\numprint{10084168} \n^2
		-\numprint{61017200} \n
		-\numprint{117867424} 
	) \mzv{3}
\\&\quad
	-45 (
		29 \n^5
		+\numprint{22644} \n^4
		-\numprint{3225892} \n^3
		-\numprint{88418816} \n^2
		-\numprint{536820560} \n
		-\numprint{897712992} 
	) 
\Big] \frac{g^7}{\numprint{41990400}}
\\&
+\bigo{g^8}
\end{split}\end{equation}
with numeric values given in table~\ref{tab:beta-numeric}.
The expansions \eqref{eq:gamma-phi}, \eqref{eq:gamma-m^2} and \eqref{eq:betaN} are available in computer readable form in the attached files (see \appendixname~\ref{sec:files}).
\begin{table}
	\centering%
	\caption{Numerical values for the $6$-loop beta function.}%
	\label{tab:beta-numeric}%
	\begin{tabular}{cc}
	\toprule
		$\quad\n\quad$ & $\beta^{\MS}(g)$   \\
	\midrule
		$0$ & $-2\eps g + 2.667 g^2 - 4.667 g^3 + 25.46 g^4-200.9 g^5+ 2004 g^6-23315 g^7+\bigo{g^8}$ \\
		$1$ & $-2\eps g+3.000 g^2-5.667 g^3+32.55 g^4-271.6g^5+2849 g^6-34776 g^7+\bigo{g^8}$ \\
		$2$ & $-2 \eps g+3.333 g^2-6.667 g^3+39.95 g^4-350.5 g^5+3845g^6-48999 g^7+\bigo{g^8}$ \\
		$3$ & $-2 \eps g+3.667 g^2-7.667 g^3+47.65 g^4-437.6 g^5+4999 g^6-66243 g^7+\bigo{g^8}$ \\
		$4$ & $-2 \eps g+4.000 g^2-8.667 g^3+55.66 g^4-533.0 g^5+6318 g^6-86768 g^7+\bigo{g^8}$ \\
	\bottomrule
	\end{tabular}%
\end{table}

\subsection{Checks}
\label{sec:checks}

Our computation exactly reproduced the full $5$-loop results \cite{ChetyrkinKataevTkachov:5loopPhi4,ChetyrkinGorishnyLarinTkachov:5loopPhi4,Kazakov:MethodOfUniqueness,KNFCL:5loopPhi4} and the 6-loop field anomalous dimension \cite{BatkovichKompanietsChetyrkin:6loop}, which in turn is consistent with the first three terms in the large $\n$ expansion of the critical exponent $\eta$ from \cite{VasilevPismakHonkonen:SimpleMethod,VasilevPismakHonkonen:largeNeta,VasilevPismakHonkonen:1n3}.
The leading and subleading terms in the large $\n$ expansion of the $6$-loop beta function, computed almost $20$ years ago in \cite{Gracey:LargeNf,BroadhurstGraceyKreimer:PositiveKnots}, provided a further successful check.

In addition, we confirmed David Broadhurst's results for the {\MS}-renormalized 5-loop propagator, obtained in $1993$. In fact, in \cite{Broadhurst:WithoutSubtractions} he obtained the $\eps$-expansions of all $5$-loop propagator integrals with $\leq 5$ loops, including the finite parts.\footnote{%
	We are very thankful to David Broadhurst for making his notes available to us.
}
These results agree perfectly with our $\eps$-expansions for those graphs.

A particularly strong check of our results is due to Oliver Schnetz \cite{Schnetz:NumbersAndFunctions}, who computed $\beta$ and $\gamma_{\field}$ to $6$ loops for the case $\n=1$, using completely different techniques including single-valued integration and graphical functions \cite{Schnetz:GraphicalFunctions}.

Also, we initially computed all necessary integrals with a combination of:
\begin{itemize}
	\item integration by parts (IBP) \cite{ChetyrkinTkachov:IBP},
	\item $\eps$-expansions of $4$-loop massless propagators \cite{BaikovChetyrkin:FourLoopPropagatorsAlgebraic,SmirnovTentyukov:FourLoopPropagatorsNumeric,LeeSmirnov:FourLoopPropagatorsWeightTwelve},
	\item infrared rearrangement (IRR) extended by the $\Rstar$-operation \cite{Chetyrkin:RRR,ChetyrkinGorishnyLarinTkachov:Analytical5loop,ChetyrkinTkachov:InfraredR,ChetyrkinSmirnov:Rcorrected,BatkovichKompaniets:6loop-Rstar} and
	\item parametric integration, using hyperlogarithms, of primitive (free of subdivergences) linear combinations of graphs; examples of this technique are given in \cite[graphs $M$, $M_{3,5}$ and $M_{5,1}$]{Panzer:MasslessPropagators} and \cite[section~5.3.2]{Panzer:PhD}.
\end{itemize}
Together with our more recent strategy \cite{KompanietsPanzer:LL2016} of parametric integration with the one-scale scheme, we have in fact computed all diagrams with at least two different exact methods ourselves.
In addition, the most complicated diagrams were also checked numerically using sector decomposition \cite{BinothHeinrich:SectorDecomposition} to at least 3 significant digits, using a computer program by the first author.
We furthermore cross-checked our generation of the Feynman graphs and their symmetry factors with {\GraphState} by comparison with the output of the program {\feyngen} from \cite{Borinsky:feyngen}. Our numbers of $2$- and $4$-point graphs in table~\ref{tab:graph-counts} agree with the lists in \cite{NickelMeironBaker:Compilation24} (after discarding the tadpole contributions).
The $O(\n)$ group factors were confirmed independently by evaluating \eqref{eq:group-factor} with a simple {\FORM} \cite{Vermaseren:NewFORM} program.

Also, we verified that, after explicitly expanding $\beta(g,\eps)=-2\eps/[\partial_g \log (gZ_g)]$ from \eqref{eq:beta-def} and $\gamma_i(g)=\beta(g) \partial_g \log  Z_i(g)$ from \eqref{eq:anom-dims} as series in $g$, all poles in $\eps$ cancel and the results indeed coincide with the final expressions in \eqref{eq:beta-def} and \eqref{eq:anom-dims} in terms of the residues $Z_{i,1}$. This shows that all higher order poles of the $Z$-factors are consistent with the first order poles as dictated by the renormalization group.

\begin{table}
	\caption{Our $6$-loop coefficients of the RG functions, compared to asymptotic Pad\'{e}-approximant predictions from \cite{EllisKarlinerSamuel:betaQCDprediction,ChishtieEliasSteele:MassivePhi4}. The errors are given as $(\text{APAP}-\text{exact})/\text{exact}$.}%
	\label{tab:APAP}%
	\centering%
	\begin{tabular}{rrrrrrr}
	\toprule
		$\n$ & 0 & 1 & 2 & 3 & 4 & 5 \\
	\midrule
		$\beta^{\MS}_7$         & 23315 & 34776 & 48999 & 66243 & 86768 & 110840 \\
		APAP \cite{EllisKarlinerSamuel:betaQCDprediction} & 
					23656 & 35374 & 49916 & 67604 & 88660 & \\
		error                 & 1.46\% & 1.72\% & 1.87\% & 2.06 \% & 2.18\% & \\
		APAP \cite{ChishtieEliasSteele:MassivePhi4} &
					& 35233 & 49381 & 66426 & 86636 & 110292 \\
		error                 & & 1.31\% & 0.78\% & 0.28\% & -0.15\%& -0.50\% \\
	\midrule
		$\gamma^{\MS}_{m^2}   $& & 1355 & 2434 & 3950 & 5983 & 8618    \\
		APAP \cite{ChishtieEliasSteele:MassivePhi4}              & & 1478 & 2740 & 4803 & 9476 &-3374    \\
		error                 & & 9\% & 13\%& 22\%& 58\%& -139\% \\ 
	\midrule
		$\gamma^{\MS}_{\field}$& & 14.4 & 24.7& 38.4 & 55.9 & 77.5  \\
		APAP \cite{ChishtieEliasSteele:MassivePhi4} &
					 & 11.2 & 20.7 & 35.0& 56.2 & 87.3 \\
		error                  & & -22.0\% & -16.3\% &-8.9\%&0.5\% & 12.6\% \\
	\bottomrule
	\end{tabular}
\end{table}

Finally, we find our results for the $6$-loop coefficients of the RG functions to be in good agreement with various past predictions based on the $\leq 5$-loop coefficients. For example, $\beta^{\MS}_7 \approx 34400$ \cite{ChetyrkinGorishnyLarinTkachov:5loopPhi4} and $\beta^{\MS}_7 \approx 34393$ \cite{Kompaniets:ACAT2016} for the case $\n=1$ are very close to our value $34776$, where $\beta^{\MS} = \sum_k (-g)^k \beta^{\MS}_k$.
Even more interestingly, \emph{asymptotic Pad\'{e}-approximant predictions} (APAPs) were provided for various values of $\n$ in \cite{ChishtieEliasSteele:MassivePhi4,EllisKarlinerSamuel:betaQCDprediction}, as summarized in table~\ref{tab:APAP}. 
In the case of the beta function, we see indeed only very small deviations ($\lesssim 2\%$) from our exact results.
It seems plausible to expect that the APAP method might provide potentially even more accurate predictions for the $7$ loop coefficient.

For the anomalous dimensions, the APAP forecasts are still reasonable though significantly less precise, as was already noted and discussed in \cite{ChishtieEliasSteele:MassivePhi4} at the five loop level.\footnote{\label{foot:APAP-errors}%
	The outlyingly large deviations in $\gamma_{m^2}$ for $\n=4$ and $\n=5$ at six loops are caused by a pole of the Pad\'{e} approximant at $\n \approx 4.7$; the agreement becomes much better again for $\n=6$ and $\n=7$.
	We are grateful to the authors of \cite{ChishtieEliasSteele:MassivePhi4} for their correspondence and investigation of this matter. In conclusion, the discrepancy at $\n=5$ is very likely an artifact of the Pad\'{e} method than an indicator of an error in our result for $\gamma_m^2$.
}
In this context let us point out that a qualitatively similar situation occurs for predictions based on the conformal Borel technique: the $5$- and $6$-loop predictions \cite{Kompaniets:ACAT2016} (at $\n=1$) for the beta function are good within $1\%$, whereas the error of the predictions \cite{BatkovichKompanietsChetyrkin:6loop} for $\eta$ increases from $0.5\%$ for $6$ loops to $2.5\%$ for $7$ loops (according to the $7$-loop result for $\eta$ given in \cite{Schnetz:NumbersAndFunctions}).

In the following, we discuss the dependence of the RG function coefficients on the perturbative order and on the internal degrees of freedom. We will see that our results are consistent with the expected behaviour.

\subsection{Asymptotics}
\label{sec:asymptotics}

It has long been known that the renormalization group functions are asymptotic series in the coupling $g$, with factorially growing coefficients \cite{Lipatov:DivergencePT,BrezinGuillouZinnJustin:phi2N}. For the minimal subtraction scheme, the precise leading asymptotic behaviour was first computed in \cite{McKaneWallaceBonfim} using $4-2\eps$ dimensional instantons \cite{McKaneWallace:Instanton}. Namely, if we denote the coefficients of the beta function by $\beta^{\MS}(g) = \sum_{k} \beta^{\MS}_{k} (-g)^k$, then
\begin{equation}
	\beta^{\MS}_{k}
	\sim
	\overline{\beta}_{k}
	\defas k! \cdot k^{3+\n/2} \cdot C_{\beta}
	\quad\text{as $k \rightarrow \infty$}
	\label{eq:beta-asymptotics}%
\end{equation}
where $C_{\beta}$ is a constant that only depends on the number $\n$ of field components:
\begin{equation}
	C_{\beta} = 
	\frac{36\cdot 3^{(\n+1)/2}}{\pi \Gamma(2+\n/2) \GK^{2\n+4}} \exp\left[-\frac{3}{2}-\frac{n+8}{3}\left(\EM+\frac{3}{4}\right)\right]
	.
	\label{eq:beta-asymptotics-constant}%
\end{equation}
In this expression, $\EM=-\Gamma'(1) \approx 0.577$ denotes the Euler-Mascheroni constant and $\GK \approx 1.282$ is the Glaisher-Kinkelin constant defined by
\begin{equation}
	\ln \GK 
	= \frac{1}{12} - \zeta'(-1)
	= \frac{1}{12} \left( \EM + \ln (2\pi) - \frac{\zeta'(2)}{\zeta(2)} \right)
	.
	\label{eq:Kinkelin}%
\end{equation}
\begin{figure}
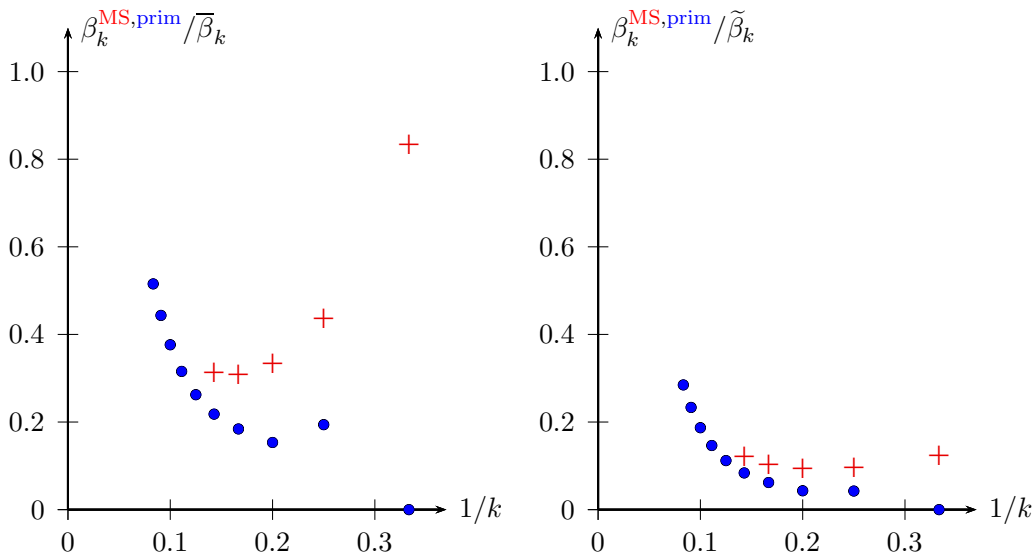

	\centering%
	\includegraphics[width=0.4\textwidth]{{{plots/beta.fac}}}
	\quad
	\includegraphics[width=0.4\textwidth]{{{plots/beta.gamma}}}
	\caption{%
		These plots demonstrate that the perturbative coefficients of the $\beta$ function ($\n=1$) are rather far from their asymptotic values:
		crosses show the known coefficients $\beta^{\MS}_k$ ($k \leq 7$) and circles indicate the estimates for the primitive contributions $\beta^{\prim}_k$ up to loop order $11$ (see table~\ref{tab:primitive-estimates}).
		They are normalized by the predictions from the asymptotic formulas \eqref{eq:beta-asymptotics} (left plot) and \eqref{eq:beta-asymptotics-gamma} (right plot). The abscissa is $1/k$, such that in the limit $k\rightarrow \infty$ the points should approach $1.0$ on the vertical axis. 
		Note that there are no primitives with two loops, resulting in the point $(1/3,0)$.
	}%
	\label{fig:beta-asymptotics}%
\end{figure}%
It is well-known that the perturbative coefficients reach their asymptotics rather slowly in the $\field^4$ model \cite{KazakovTarasovShirkov:AnalyticContinuation,DittesKubyshinTarasov:4loop}, which is illustrated in figure~\ref{fig:beta-asymptotics} and table~\ref{tab:primitive-ratios}. We see that, even at $6$ loops, the ratio $\beta^{\MS}_7/\overline{\beta}_7 \approx 0.31$ is still far from one. 
One also should bear in mind that, at such low perturbative orders, this kind of ratio depends significantly on the function used to model the asymptotic behaviour \cite[Fig.~4]{KazakovTarasovShirkov:AnalyticContinuation}. For illustration, we include in figure~\ref{fig:beta-asymptotics} a comparison of $\beta^{\MS}_{k}$ to 
\begin{equation}
	\widetilde{\beta}_{k}
	\defas \Gamma(k+4+\n/2) \cdot C_{\beta}
	,
	\label{eq:beta-asymptotics-gamma}%
\end{equation}
which is another reasonable function with the same asymptotics as \eqref{eq:beta-asymptotics}.
\begin{table}
	\caption{%
		The upper part of the table shows the $\ell$-loop coefficients $\beta^{\MS}_{\ell+1}$ of the $\field^4$ beta function $\beta^{\MS} = \sum_k \beta^{\MS}_k (-g)^k$ (with $\n=1$) and their ratios to the asymptotic formulas \eqref{eq:beta-asymptotics} and \eqref{eq:beta-asymptotics-gamma}.
		In the lower two rows we show the contribution $\beta^{\prim}_{\ell+1}$ containing only the primitive graphs, and their proportion of the full $\ell$-loop coefficient.
	}%
	\label{tab:primitive-ratios}%
	\centering
	\begin{tabular}{rrrrrrr}
	\toprule%
	loop order $\ell$        
		&   1 &    2 &    3 &   4 &    5 &     6 \\
	\midrule%
	$\beta^{\MS}_{\ell+1}/\overline{\beta}_{\ell+1}$ in \%
		& 548 & 83.5 & 43.8 &33.5 & 30.9 &  31.4 \\
	$\beta^{\MS}_{\ell+1}/\widetilde{\beta}_{\ell+1}$ in \%
		&43.1 & 12.5 &  9.58& 9.41& 10.4 &  12.1 \\
	$\beta^{\MS}_{\ell+1}$   
		&   3 & 5.67 & 32.5 & 272 & 2849 & 34776 \\
	\midrule
	$\beta^{\prim}_{\ell+1}$ 
		&   3 &    0 & 14.4 & 124 & 1698 & 24130 \\
	$\beta^{\prim}_{\ell+1}/\beta^{\MS}_{\ell+1}$ in \% 
		& 100 &    0 & 44.3 &45.8 & 59.6 &  69.4 \\
	\bottomrule%
	\end{tabular}
\end{table}

In order to probe the asymptotic regime further, we studied the \emph{primitive} contributions $\beta^{\prim}_k$ to $\beta^{\MS}_k$, that is, the sum of the contributions of all $4$-point $\field^4$ graphs that are free of subdivergences.
These dominate the leading asymptotics of $\beta^{\MS}_k$ as $k \rightarrow \infty$, according to \cite[page~1865]{McKaneWallaceBonfim}:\footnote{%
	See also \cite[page~15]{BrezinGuillouZinnJustin:phi2N}: ``Finally, the leading diagrams at order $K$ give a single power of $\ln \Lambda$, they are those which do not involve any divergent subgraph; i.e., they are the completely irreducible diagrams.''
}
\begin{quote}
	In the context of high-order estimates in the perturbative series, we interpret the extra pole in $\eps$ as the one produced by the totally irreducible diagrams at high orders. These diagrams are known to be the dominant ones at $K$th order for $K$ large for $d < 4$ and moreover diverge only like $1/\eps$.
\end{quote}
Indeed, the last row of table~\ref{tab:primitive-ratios} shows how the primitive graphs become more and more relevant as the loop number increases.\footnote{%
	Such a comparison was already made in evaluation of the four loop calculation, see \cite{KazakovTarasovShirkov:AnalyticContinuation}.
}
At $6$ loops, the primitive graphs $\beta^{\prim}_7$ already constitute $69\%$ of the coefficient $\beta^{\MS}_7$.
The primitive contributions at $7$ loops are known exactly \cite{PanzerSchnetz:Phi4Coaction} and included in figure~\ref{fig:beta-asymptotics}.

Furthermore, we were able to obtain accurate numeric estimates for all primitive graphs with up to $11$ loops, using a new method (based on the so-called \emph{Hepp bound}) recently introduced by the second author. A brief sketch of this technique is provided in \appendixname~\ref{sec:Hepp-bound}.
We see in figure~\ref{fig:beta-asymptotics} that even at $11$ loops, the primitive contributions (which are expected to be close to the full $\beta^{\MS}_{12}$) reach merely about half of the value predicted by the asymptotic formula~\eqref{eq:beta-asymptotics}. Details are given in table~\ref{tab:primitive-estimates}.

We conclude that it is not clear how the knowledge of the leading asymptotic behavior of the perturbative coefficients might be used to accurately predict perturbative coefficients at higher orders.
It is thus interesting to note that, in principle, corrections to the leading asymptotic behaviour of the form
\begin{equation}
	\beta^{\MS}_k \sim k!\cdot k^{3+\n/2} \cdot C_{\beta}
	\left( 1 + \sum_{j=1}^{\infty} \frac{a_j}{k^j} \right)
	\quad \text{as $k\rightarrow \infty$}
	\label{eq:beta-asymptotics-corrections}%
\end{equation}
can indeed be calculated. In fact, the first correction $a_1$ was computed long ago in \cite{Kubyshin:Corrections} for the MOM scheme and much more recently in \cite{KomarovaNalimov:FirstCorrectionO(N)} for the {\MS} scheme, using a method developed in \cite{KomarovaNalimov:HigherOrdersO(N)}.
Unfortunately, these results need to be adjusted%
\footnote{%
	In particular, in both of these papers, the value of $U_{\text{reg}}$ needs to be corrected to $\frac{\n+8}{6} (\EM+5/3+\ln\pi)$, as was kindly pointed out to us by M.~Nalimov. Note that, with this correction, the result for the leading asymptotics computed in \cite{KomarovaNalimov:FirstCorrectionO(N)} coincides with \eqref{eq:beta-asymptotics} and \eqref{eq:beta-asymptotics-constant} from \cite{McKaneWallaceBonfim}.
}
and we therefore cannot presently discuss if the term $a_1/k$ narrows the gap between the exactly known low-order coefficients $\beta^{\MS}_k$ and the asymptotic predictions. This correction could also inflate the gap, because the expansion \eqref{eq:beta-asymptotics-corrections} is itself only asymptotic \cite{Suslov:HigherOrderLipatov,LobaskinSuslov:HigherLipatov} and as such only guarantees an improved fit for very large perturbative orders $k$.

In particular, we like to stress that the idea to truncate \eqref{eq:beta-asymptotics-corrections}, fit the coefficients $a_j$ to make this polynomial in $1/k$ match the known low order perturbative terms and then use this polynomial to predict (extrapolate) higher order perturbative coefficients, is not justified.\footnote{%
	Because \eqref{eq:beta-asymptotics-corrections} is asymptotic, it should be expected that longer truncations with the correct values of $a_j$ yield increasingly worse predictions for the low order perturbative coefficients. Hence, forcing these to be matched requires the fitted $a_j$ to deviate from their exact values, so there is no reason why those should result in good predictions for higher orders $k$.
}
For a detailed discussion and criticism of such a procedure, see~\cite{KazakovPopov:QMandQFT} and \cite{KazakovPopov:GellMannLow}.

\subsection{Dependence on $\n$}
\label{sec:n-dependence}

\begin{figure}
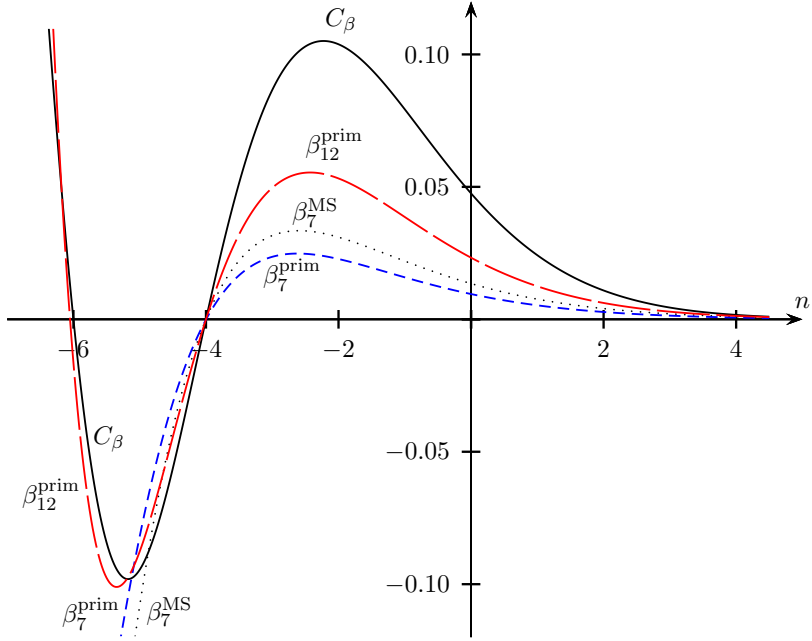

	\centering%
	\includegraphics[width=0.65\textwidth]{{{plots/beta.n-dependence}}}%
	\caption{%
		Dependence of the coefficients of the beta function on $\n$: The dashed curves show the primitive contributions $\beta^{\prim}_k/(k!\cdot k^{3+\n/2})$ at $6$ and $11$ loops.
		For comparison with the full beta function, our result for $\beta^{\MS}_k/(k!\cdot k^{3+\n/2})$ at $6$ loops is included as the dotted line.
		The solid line shows the limiting curve for $k\rightarrow \infty$, namely $C_{\beta}$ from \eqref{eq:beta-asymptotics-constant}.
	}%
	\label{fig:beta-n-dependence}%
\end{figure}

The coefficients $\beta^{\MS}_k=\beta^{\MS}_k(\n)$ are functions of the number $\n$ of fields, because the contribution of each Feynman graph is multiplied with a corresponding group factor \eqref{eq:group-factor-from-vacuum}.
More precisely, $\beta^{\MS}_k(\n)$ is a polynomial in $\n$ for each order $k$, as seen in \eqref{eq:betaN}.
Once normalized by the asymptotic growth $k!\cdot k^{3+\n/2}$ from \eqref{eq:beta-asymptotics}, these coefficients approach $C_{\beta}$ in the limit $k\rightarrow \infty$. Since this limit is dominated by the primitive graphs, our estimates for $\beta^{\prim}_k$ should exhibit the same behaviour.

Figure~\ref{fig:beta-n-dependence} shows that these expectations are indeed fulfilled. First, we note that the observation (from figure~\ref{fig:beta-asymptotics} at $\n=1$), that even the $11$th perturbative order is far from the asymptotic regime, extends to all $\n \gtrsim -3$.
However, all curves share a zero at $\n \approx -4$, and the primitive contributions with $11$ loops vanish also near to the next zero of $C_{\beta}$ at $\n=-6$. We note that the asymptotic coefficient $C_{\beta}$ is approximated rather well by the primitive contributions $\beta^{\prim}_{12}$ in the intermediate region where $-6<\n<-4$.

This phenomenon of zeros at certain values of $\n$ was observed in \cite{Pobylitsa:Superfast} and follows simply from the factor $\Gamma(2+\n/2)$ in the denominator of \eqref{eq:beta-asymptotics-constant}, since it implies that $C_{\beta}$, which governs the asymptotic behaviour of $\beta_k^{\MS}$ according to \eqref{eq:beta-asymptotics}, vanishes at even values of $\n \leq -4$.
\begin{table}
	\caption{The first zeros of the exactly known perturbative coefficients $\beta^{\MS}_{\ell+1}$ from \eqref{eq:betaN} as functions of $\n$, for $\ell \leq 6$ loops, and similarly for our estimates $\beta^{\prim}_{\ell+1}$ for the primitive contributions from \appendixname~\ref{sec:Hepp-bound} (up to $11$ loops).} %
	\label{tab:asymptotic-n-zeros}%
	\centering
	\begin{tabular}{@{}rrlrr}
	\toprule
	& loop order $\ell$ & first zero & second zero & third zero \\
	\midrule
	\llap{{\ldelim\{{6}{*}[{$\beta^{\MS}_{\ell+1}(\n)$}]}}
		& 1 & -8       &       & \\
		& 2 & -4.67    &       & \\
		& 3 & -4.025   & -41.4 & \\
		& 4 & -4.020   & -12.1 & 3219 \\
		& 5 & -4.0017  & -8.76 & -44.0 \\
		& 6 & -4.00044 & -7.52 & -20.0 \\
	\cline{2-5}
	\llap{{\ldelim\{{6}{*}[{$\beta^{\prim}_{\ell+1}(\n)$}]}}
		&  6 & -3.99754     & -7.22  & -35.6 \\
		&  7 & -3.99982     & -6.58  & -15.1 \\
		&  8 & -3.99994     & -6.31  & -10.8 \\
		&  9 & -3.999997    & -6.18 & -9.24 \\
		& 10 & -3.99999991  & -6.10 & -8.55 \\
		& 11 & -4.000000095 & -6.05 & -8.21 \\
	\bottomrule
	\end{tabular}
\end{table}
Indeed, the zero of $\beta^{\MS}_{\ell+1}(\n)$ that is closest to the origin converges rapidly towards $-4$ as $\ell$ increases, which was checked for $\ell \leq 5$ loops in \cite{Pobylitsa:Superfast}. In table~\ref{tab:asymptotic-n-zeros}, we see that this trend continues at $\ell=6$ loops and, with an impressive rate of convergence, the same phenomenon continues in our estimates of the primitive contributions $\beta^{\prim}_{\ell+1}$ up to $\ell=11$ loops.
Furthermore, at these higher loop orders, we can see the convergence of the next zeros to the expected values $\n=-6$ and $\n=-8$.

Note that the group factors $\GroupFactor{G}$ can be negative for such values of $\n$, and in fact only because of these opposite signs is the cancellation of $\beta^{\prim}_{\ell+1}(\n)$ possible at all. 
As a collective phenomenon, sensitive to all Feynman periods at loop order $\ell$, we thus interpret the convergence of zeros in table~\ref{tab:asymptotic-n-zeros} towards even values of $\n\leq-4$ as a strong consistency check of our exact $6$-loop results and also of our estimates for the primitive contributions up to $11$ loops.

\section{Resummation of critical exponents}
\label{sec:resummation}

From the renormalization group functions $\gamma_{\field}$, $\gamma_{m^2}$ and $\beta$ in section~\ref{sec:RG-functions}, it is straightforward to work out the $\eps$-expansions of critical exponents. We focus on $\eta$, $\nu^{-1}$ and $\omega$, for which \eqref{eq:def-eta-nu} and \eqref{eq:def-omega} yield the results shown in tables \ref{tab:eta-numeric}--\ref{tab:omega-numeric}.

\begin{table}
	\caption{Numerical $6$-loop $\eps$-expansion of the critical exponent $\eta$ in $\D=4-2\eps$ dimensions.}%
	\label{tab:eta-numeric}%
	\centering%
	\begin{tabular}{cc}
	\toprule
		$\quad\n\quad$ & $\eta(\eps)$   \\
	\midrule
		$0$ & $0.062500\eps^{2} +0.13281\eps^{3} -0.13388\eps^{4} +0.84815\eps^{5} -5.8067\eps^{6}+\bigo{\eps^{7}}$ \\
		$1$ & $0.074074\eps^{2} +0.14952\eps^{3} -0.13326\eps^{4} +0.82101\eps^{5} -5.2014\eps^{6}+\bigo{\eps^{7}}$ \\
		$2$ & $0.080000\eps^{2} +0.15200\eps^{3} -0.12630\eps^{4} +0.74269\eps^{5} -4.3921\eps^{6}+\bigo{\eps^{7}}$ \\
		$3$ & $0.082645\eps^{2} +0.14719\eps^{3} -0.11919\eps^{4} +0.65225\eps^{5} -3.6495\eps^{6}+\bigo{\eps^{7}}$ \\
		$4$ & $0.083333\eps^{2} +0.13889\eps^{3} -0.11336\eps^{4} +0.56421\eps^{5} -3.0312\eps^{6}+\bigo{\eps^{7}}$ \\
	\bottomrule
	\end{tabular}%
\end{table}%
\begin{table}
	\caption{Numerical $6$-loop $\eps$-expansion of the critical exponent $1/\nu$ in $\D=4-2\eps$ dimensions.}%
	\label{tab:invnu-numeric}%
	\centering%
	\begin{tabular}{cc}
	\toprule
		$\quad\n\quad$ & $\nu^{-1}(\eps)$   \\
	\midrule
		$0$ & $2.0000 -0.50000\eps -0.34375\eps^{2} +0.91540\eps^{3} -4.6002\eps^{4} +30.596\eps^{5} -246.77\eps^{6}+\bigo{\eps^{7}}$ \\
		$1$ & $2.0000 -0.66667\eps -0.46914\eps^{2} +0.99622\eps^{3} -4.9096\eps^{4} +30.440\eps^{5} -228.65\eps^{6}+\bigo{\eps^{7}}$ \\
		$2$ & $2.0000 -0.80000\eps -0.56000\eps^{2} +0.97949\eps^{3} -4.8756\eps^{4} +28.136\eps^{5} -198.59\eps^{6}+\bigo{\eps^{7}}$ \\
		$3$ & $2.0000 -0.90909\eps -0.62359\eps^{2} +0.92057\eps^{3} -4.6976\eps^{4} +25.278\eps^{5} -168.91\eps^{6}+\bigo{\eps^{7}}$ \\
		$4$ & $2.0000 -1.00000\eps -0.66667\eps^{2} +0.84684\eps^{3} -4.4586\eps^{4} +22.469\eps^{5} -142.96\eps^{6}+\bigo{\eps^{7}}$ \\
	\bottomrule
	\end{tabular}%
\end{table}%
\begin{table}
	\caption{Numerical $6$-loop $\eps$-expansion of the correction to scaling $\omega$ in $\D=4-2\eps$ dimensions.}%
	\label{tab:omega-numeric}%
	\centering%
	\begin{tabular}{cc}
	\toprule
		$\quad\n\quad$ & $\omega(\eps)$   \\
	\midrule
		$0$ & $2.0000\eps -2.6250\eps^{2} +14.589\eps^{3} -100.57\eps^{4} +859.94\eps^{5} -8320.5\eps^{6}+\bigo{\eps^{7}}$ \\
		$1$ & $2.0000\eps -2.5185\eps^{2} +12.946\eps^{3} -83.762\eps^{4} +663.99\eps^{5} -5959.1\eps^{6}+\bigo{\eps^{7}}$ \\
		$2$ & $2.0000\eps -2.4000\eps^{2} +11.497\eps^{3} -70.724\eps^{4} +523.96\eps^{5} -4401.7\eps^{6}+\bigo{\eps^{7}}$ \\
		$3$ & $2.0000\eps -2.2810\eps^{2} +10.263\eps^{3} -60.498\eps^{4} +421.82\eps^{5} -3341.1\eps^{6}+\bigo{\eps^{7}}$ \\
		$4$ & $2.0000\eps -2.1667\eps^{2} +9.2207\eps^{3} -52.351\eps^{4} +345.65\eps^{5} -2596.3\eps^{6}+\bigo{\eps^{7}}$ \\
	\bottomrule
	\end{tabular}%
\end{table}%

The $\eps$-expansion
$
	f(\eps) = \sum_{k=0}^{\infty} f_k (-2\eps)^k
$
of a critical exponent $f$ around $\D=4-2\eps$ dimensions is a formal power series with factorially growing coefficients
\begin{equation}
	f_k \sim C_f \cdot k! \cdot a^k \cdot k^{b_f}
	\quad\text{as}\quad
	k\rightarrow \infty
	.
	\label{eq:coefficient-asymptotics}%
\end{equation}
In fact, this leading asymptotic behaviour is completely determined by the asymptotics \eqref{eq:beta-asymptotics} of the beta function, the leading terms \eqref{eq:anom-first-order} of the perturbation series and the defining equations \eqref{eq:gcrit-first-order}, \eqref{eq:def-omega} and \eqref{eq:def-eta-nu}.%
\footnote{%
	The equations \eqref{eq:gcrit-first-order}--\eqref{eq:def-eta-nu} furthermore relate corrections to the leading asymptotics, as for example computed in \cite{KomarovaNalimov:FirstCorrectionO(N)}. If one encodes the full asymptotic expansions as generating functions, these relations can be computed elegantly in an algebraic way \cite{Borinsky:GeneratingAsymptotics}.
}
In particular, all the coefficients $C_f$ can be expressed as multiples of $C_{\beta}$ from \eqref{eq:beta-asymptotics-constant} and were all computed in \cite{McKaneWallaceBonfim}. The values of $a$ and the exponents $b_f$ were already obtained in \cite{BrezinGuillouZinnJustin:phi2N}:
\begin{equation}
	a = \frac{3}{\n+8}
	\quad\text{and}\quad
	b_f = \begin{cases}
		    3 + \n/2 & \text{for $f=\eta$,}\\
		    4 + \n/2 & \text{for $f=\nu^{-1}$ and}\\
		    5 + \n/2 & \text{for $f=\omega$.}\\
	      \end{cases}
	\label{eq:asymptotic-exponents}%
\end{equation}
In order to obtain estimates for the critical exponents in $\D=3$ dimensions, we must resum the divergent series $f(\eps)$ at $\eps=1/2$.

This problem of resummation is a huge subject, see for example the review \cite{Caliceti:UsefulAlgorithms}, and many different approaches have been put forward. Unfortunately, no consensus has so far been reached on the optimal method to resum $\eps$-expansions.
We therefore think that a careful comparison of the various methods, based on the $6$-loop perturbation series presented here, would be very valuable. In particular in view of the potential for further higher order perturbative computations in the near future, like the $7$-loop $\eps$-expansions in the $\field^4$ model \cite{Schnetz:NumbersAndFunctions}, such further insight into the resummation problem is very desirable.

However, such an extensive analysis would exceed the scope of this article and we decided to discuss only the method of Borel resummation with conformal mapping.
It would be very interesting to see how other approaches, including order-dependent mapping \cite{ZinnJustin:ODM,SeznecZinnJustin:OrderDependent}, large-coupling expansions \cite{Kleinert:ConvertingWeakStrong,KleinertFrohlinde:5loopStrong,Kleinert:7loopPhi4strong} and self-similar factor approximants \cite{YukalovYukalova:CriticalSelfSimilar}, fare with the $6$-loop perturbative input.

\subsection{Borel resummation with conformal mapping}
We will describe the method introduced first in \cite{KazakovTarasovShirkov:AnalyticContinuation} for the resummation of series in the coupling $g$ and then also applied to $\eps$-expansions \cite{VladimirovKazakovTarasov:Calculation}. This technique has a successful history in the resummation of critical exponents \cite{LeGuillouZinnJustin:CriticalFromField,GuillouZinnJustin:Accurate,GuidaZinnJustin:CriticalON,GorishnyLarinTkachov:O(eps5)} and is explained in detail for example in \cite[chapter~16]{KleinertSchulteFrohlinde:CriticalPhi4}.

To begin with, we denote the \emph{Borel transform} of $f=\sum_{k=0}^{\infty} f_k (-2\eps)^k$ as
\begin{equation}
	\Borel[b]{f}(x)
	\defas \sum_{k=0}^{\infty}
	\frac{f_k}{\Gamma(k+b+1)} (-x)^k
	.
	\label{eq:borel}%
\end{equation}

According to \eqref{eq:coefficient-asymptotics}, it defines an analytic function in the domain $\abs{x} < 1/a$  with a singularity at $x=-1/a$ of the form $(1+ax)^{b-b_f-1}$.
We assume%
\footnote{%
	While the Borel summability with respect to the coupling $g$ has been established in the fixed dimensions $\D=2$ \cite{EckmannMagnenSeneor:Borel} and $\D=3$ \cite{MagnenSeneor:phi4_3}, it remains an open question for the $\eps$-expansion.
}
that $f$ is Borel summable, which means that $\Borel[b]{f}(x)$ admits an analytic continuation to the positive real axis $x>0$ and also includes that the \emph{Borel sum}
\begin{equation}
	\tilde{f}(\eps)
	\defas \int_0^{\infty} t^{b} e^{-t} \Borel[b]{f}(2\eps t) \ \dd t
	\label{eq:inverse-borel}%
\end{equation}
converges and gives the correct value of the critical exponent at $\eps=1/2$, which is our case of interest ($\D=3$). By construction, this Borel sum $\tilde{f}(\eps)$ has the perturbative expansion $f(\eps)$ as required.
In order to compute the integral \eqref{eq:inverse-borel}, we must analytically continue the Borel transform $\Borel[b]{f}(x)$ from the circle $\abs{x}<1/a$ of convergence to the positive real line. This is achieved with the conformal transformation
\begin{equation}
	w(x) = \frac{\sqrt{1+ax}-1}{\sqrt{1+ax}+1}
	\quad\text{with inverse}\quad
	x(w) = \frac{4w}{a(1-w)^2}
	,
	\label{eq:conformal-mapping}%
\end{equation}
as it maps the integration domain $x\in(0,\infty)$ to the interval $w\in (0,1)$. Assuming that all singularities of $\Borel[b]{f}(x)$ lie on the cut $(-\infty,-1/a]$, which is mapped onto the unit circle $\abs{w}=1$, the expansion of the Borel transform $\Borel[b]{f}(x(w))$ into a series in $w$ converges in the full integration domain and thus provides the sought-after analytic continuation.

Because we only know the first few expansion coefficients $f_k$ with $k\leq 6$, we can only approximate the Borel transform.
Following \cite{KazakovTarasovShirkov:AnalyticContinuation,VladimirovKazakovTarasov:Calculation}, we introduce a second parameter $\lambda$ and the truncation order $\ell$ to write it as
\begin{equation}
	\Borel[b]{f}(x)
	\approx
	\Borel[b,\lambda,\ell]{f}(x)
	\defas \left( \frac{ax}{w(x)} \right)^{\lambda}
	\sum_{k=0}^{\ell} B^{b,\lambda}_{f,k} \left[w(x)\right]^k
	.
	\label{eq:borel-lambda-ansatz}%
\end{equation}
The coefficients $B_{f,k}^{b,\lambda}$ are functions of $b$ and $\lambda$, determined by matching the coefficients of $x^0$ through $x^{\ell}$ in the expansion of $\Borel[b,\lambda,\ell]{f}(x)$ with the perturbative constraints \eqref{eq:borel}. This ensures that $\Borel[b]{f}(x)$ is approximated well for small values of $x$.
Crucially, the parameter $\lambda$ allows us to also adjust the growth $\Borel[b,\lambda,L]{f}(x) \sim x^{\lambda}$ for large $x$ to better match the behaviour of the actual Borel transform. 
Without this degree of freedom, our approximations $\Borel[b,0,\ell]{f}(x) \rightarrow \sum_{k=0}^{\ell} B_{f,k}^{b,0}$ would always approach a constant value at $x\rightarrow \infty$. 
This appears to model the Borel transform only very poorly, and a careful choice of $\lambda$ is essential to significantly improve the quality of the resummation \cite{VladimirovKazakovTarasov:Calculation,GorishnyLarinTkachov:O(eps5),Kompaniets:ACAT2016}.

Our approximate result for the Borel sum~\eqref{eq:inverse-borel} is then given by
\begin{equation}
	\tilde{f}(\eps)
	\approx
	\tilde{f}_{\ell}^{b,\lambda}(\eps)
	\defas \int_0^{\infty} t^{b} e^{-t} \Borel[b,\lambda,\ell]{f}(2\eps t) \ \dd t
	.
	\label{eq:borel-inverse-truncated}%
\end{equation}

Finally, a third parameter was introduced in \cite{GuillouZinnJustin:Accurate} to improve the results even further. Namely, we consider a \emph{homographic transformation}
\begin{equation}
	\eps  = h_q(\eps') \defas \frac{\eps'}{1+q\eps'}
	\quad\text{with inverse}\quad
	\eps' = h_q^{-1}(\eps) = \frac{\eps}{1-q\eps}
	\label{eq:homographic}%
\end{equation}
to re-expand the original $\eps$-expansions as series in $\eps'$ (for $q=0$, nothing changes).
We then proceed as above, taking this new series as input:
\begin{equation}
	\tilde{f}(\eps)
	\approx
	\tilde{f}_{\ell}^{b,\lambda,q}(\eps)
	\defas \int_0^{\infty} t^{b} e^{-t} \Borel[b,\lambda,\ell]{f\circ h_q}\left(\frac{2\eps t}{1-q\eps}\right) \ \dd t
	.
	\label{eq:borel-inverse-homographic}%
\end{equation}
The motivation for \eqref{eq:homographic} is that it allows us to map potential singularities of critical exponents as functions of $\eps$ further away from the point $\eps=1/2$, in order to diminish their possibly detrimental influence on the resummation \cite{GuillouZinnJustin:Accurate}.

In closing, let us stress that there are several aspects in which this basic scheme might be adjusted. For example, we could choose other conformal mappings than \eqref{eq:conformal-mapping}, replace \eqref{eq:homographic} by a different transformation and instead of \eqref{eq:borel-lambda-ansatz} we could use another class of functions to approximate the Borel transform.%
\footnote{For example, hypergeometric functions were proposed in \cite{MeraPedersenNikolic:Nonperturbative} and \cite[chapter~19]{KleinertSchulteFrohlinde:CriticalPhi4}. Furthermore, Pad\'{e} approximants have been suggested as a replacement for the Taylor series \eqref{eq:borel-lambda-ansatz}; see \cite{JentschuraSoff:ImprovedConformal,GaddahJwan:ConformalBorel}.}

\subsection{Dependence on resummation parameters}
\label{sec:parameter-dependence}

Our resummation procedure is formulated in terms of three parameters: $b$, $\lambda$ and $q$. 
If we knew the full perturbation series of a critical exponent $f(\eps)$, the resummed value $\tilde{f}(\eps)=\lim_{\ell\rightarrow \infty} \tilde{f}^{b,\lambda,q}_{\ell}(\eps)$ would not depend on these choices.
But since we only have the first few terms of the $\eps$-expansions at hand, we are forced to consider the truncations $\tilde{f}^{b,\lambda,q}_{\ell}(\eps)$ with $\ell \leq 6$. 
These do depend on the parameters, which therefore have to be chosen carefully.

Let us first comment on $b$. The asymptotic behaviour of the coefficients of our approximations of the Borel transform, $\Borel[b,\lambda,\ell]{f}(x)$ from \eqref{eq:borel-lambda-ansatz}, is given by
\begin{equation}
\begin{split}
	\left( \frac{ax}{w(x)} \right)^\lambda \left[ w(x) \right]^p
	&= 4^{\lambda} \left( \frac{ax}{4} \right)^p
	\sum_{k=0}^{\infty} \binom{2(k-\lambda+p)}{k} \frac{\lambda-p}{\lambda-p-k} \left( -\frac{ax}{4} \right)^k
	\\
	&=(-1)^{p}(p-\lambda) \sum_{k=p}^{\infty} \frac{(-ax)^k}{\sqrt{\pi k^3}}\left( 1+\bigo{\frac{1}{k}} \right)
	.
\end{split}
\label{eq:conformal-asymptotic-expansion}%
\end{equation}
It was noted in \cite{KazakovTarasovShirkov:AnalyticContinuation} that we can therefore match the leading asymptotics \eqref{eq:coefficient-asymptotics} of the perturbation series and our model for the Borel transform by setting $b=b_f+3/2$, according to \eqref{eq:borel} and $\Gamma(k+b+1) \sim k! \cdot k^b$. 
This fixed value was indeed used in \cite{KazakovTarasovShirkov:AnalyticContinuation,VladimirovKazakovTarasov:Calculation,GorishnyLarinTkachov:O(eps5),BatkovichKompanietsChetyrkin:6loop,Kompaniets:ACAT2016}, with the idea that it incorporates the contributions from very high order perturbation theory.
We do not follow this strategy, for the following reasons:
\begin{enumerate}
	\item 
	For an exact resummation of the high order contributions, we would actually also have to enforce a precise matching of the constant\footnote{This was pointed out in \cite{KazakovTarasovShirkov:AnalyticContinuation}. We note that in this reference, the overall sign in equation~(14) seems wrong, and in equations~(13) and (14), $a_0$ should not appear, according to our \eqref{eq:conformal-asymptotic-expansion}.}
\begin{equation*}
	C_f 
	\stackrel{!}{=} 
	\frac{1}{\sqrt{\pi}}
	\sum_{p=0}^L (-1)^p (p-\lambda) B^{b,\lambda}_{f,p}
	.
\end{equation*}

	\item In section~\ref{sec:asymptotics} we saw that even six loops remain far away from the asymptotic regime (figure~\ref{fig:beta-asymptotics}). The contribution of the resummed asymptotic higher orders is thus likely outweighed by the deviation of the $7$-loop contribution from its asymptotic estimate \eqref{eq:coefficient-asymptotics}.

	\item Variation of the parameter $b$, as first suggested in \cite{LeGuillouZinnJustin:n3D}, can improve the resummation and also provides a hint towards the uncertainty of the result.
\end{enumerate}
Instead, let us investigate how the resummation depends on $b$.
According to \eqref{eq:borel}, the Taylor coefficients of the Borel transform become smaller with increasing $b$. 
Since the non-truncated Borel sum does not depend on $b$, this implies that the dominant contribution to the integral~\eqref{eq:inverse-borel} must come from larger values of $x$. Hence, with increasing $b$, our model for the large-$x$ behaviour of the Borel transform (encoded in the parameter $\lambda$) should become more relevant.

\begin{figure}
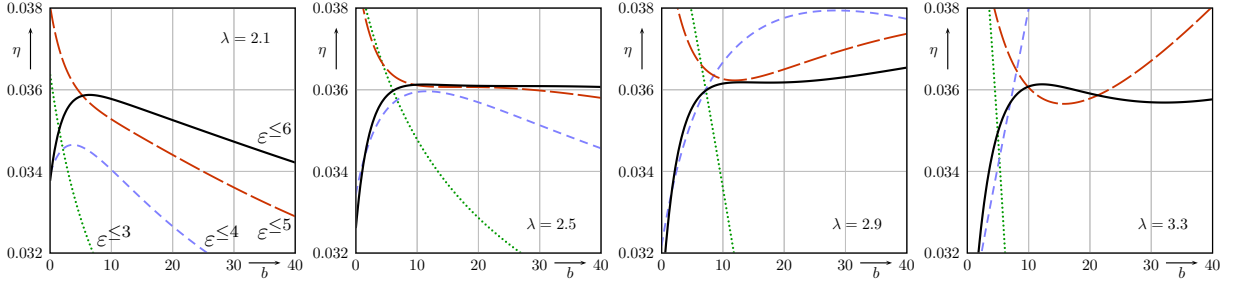

	\centering%
	\includegraphics[width=0.24\textwidth]{{{plots/eta3n1q0.2lambda2.1}}}~%
	\includegraphics[width=0.24\textwidth]{{{plots/eta3n1q0.20lambda2.5}}}~%
	\includegraphics[width=0.24\textwidth]{{{plots/eta3n1q0.2lambda2.9}}}~%
	\includegraphics[width=0.24\textwidth]{{{plots/eta3n1q0.2lambda3.3}}}%
	\caption{Dependence of the resummations $\tilde{\eta}^{b,\lambda,q}_{\ell}$ of $\eta$ on $b$ for different values of $\lambda$, at $\n=1$ in $\D=3$ dimensions ($\eps=0.5$) with $q=0.2$. The loop order $\ell$ is indicated by the label $\eps^{\leq \ell}$.}%
	\label{fig:b-plots-of-lambda}%
\end{figure}

Figure~\ref{fig:b-plots-of-lambda} demonstrates this behaviour, by plotting the resummations $\tilde{\eta}^{b,\lambda,q}_{\ell}(1/2)$ of the critical exponent $\eta$ (with various truncation orders $\ell$ of the $\eps$-expansion) as a function of $b$ for several values of $\lambda$.
First note that, as expected, in each case the dependence on $b$ decreases if we take more terms of the $\eps$-expansion into account (it will disappear completely only in the limit $\ell \rightarrow \infty$).
Furthermore, the choice of $\lambda$ has a strong influence on the $b$-dependence of the resummation. 

In fact, for most $\lambda$'s, the dependence on $b$ is very strong. 
However, we find that for a rather small range of $\lambda$, the $5$- and $6$-loop resummations become almost insensitive to $b$ as illustrated by the plateau in the second plot of figure~\ref{fig:b-plots-of-lambda}.
This stability (with respect to $b$) deteriorates quickly if we resum fewer terms of the $\eps$-expansion.
As expected from our discussion earlier, we also see that for larger values of $b$, the resummation depends more strongly on the tuning of $\lambda$.
Finally note that with \eqref{eq:asymptotic-exponents}, the value $b=b_{\eta}+3/2=5$ chosen in \cite{VladimirovKazakovTarasov:Calculation,GorishnyLarinTkachov:O(eps5)} seems too small and misses the plateaus.

In conclusion, we take the sensitivity with respect to $b$ both as an indicator of the uncertainty of the resummation and as a criterion to choose $\lambda$ and $q$. This is a common approach in general, and the existence of wide plateaus in this case was for example pointed out in \cite{MudrovVarnashev:ModifiedBorel}.\footnote{%
	In \cite{MudrovVarnashev:ModifiedBorel,DelamotteDudkaHolovatchMouhanna:Frustrated}, the same approach was used to optimize the value for $a$ in \eqref{eq:conformal-mapping}. It was observed that the dependence on $a$ is very weak and that the best choices of $a$ lie very close to the value from \eqref{eq:asymptotic-exponents} in the asymptotic growth \eqref{eq:coefficient-asymptotics}. We therefore keep $a$ fixed at this value.
}
The power of studying the sensitivity with respect to the resummation parameters was also demonstrated in \cite{DelamotteDudkaHolovatchMouhanna:Frustrated}, and indeed we will also apply this criterion with respect to variations of $\lambda$.

\begin{figure}
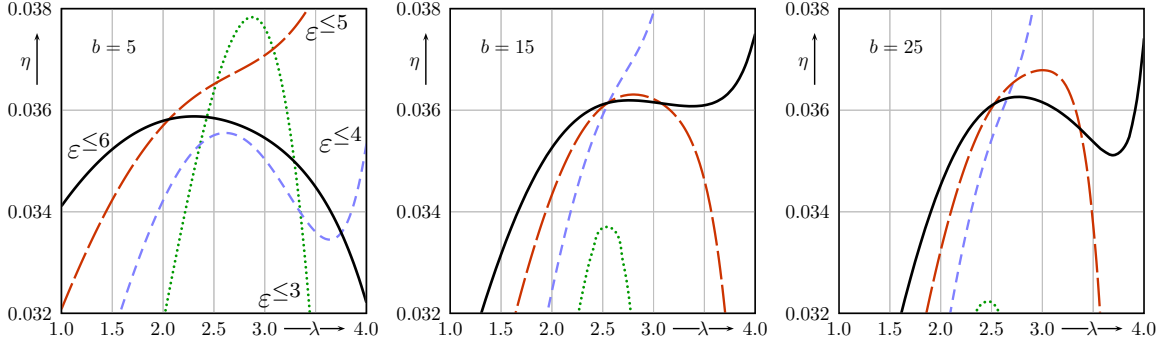

	\centering%
	\includegraphics[width=0.3\textwidth]{{{plots/eta3n1q0.2b5}}}~
	\includegraphics[width=0.3\textwidth]{{{plots/eta3n1q0.2b15}}}~
	\includegraphics[width=0.3\textwidth]{{{plots/eta3n1q0.2b25}}}
	\caption{Dependence of the resummation of $\eta$ on $\lambda$ for different values of $b$, for $\n=1$ in $\D=3$ dimensions with $q=0.2$. The loop order $\ell$ is indicated by the label $\eps^{\leq \ell}$.}%
	\label{fig:lambda-plots-of-b}%
\end{figure}

In figure~\ref{fig:lambda-plots-of-b} we see the dependence of the resummation on $\lambda$ for various $b$. As we expect from figure~\ref{fig:b-plots-of-lambda}, very small values like $b=5$ give a very unstable picture. For suitable larger values like $b=15$ we find $\lambda$-intervals where the six loop resummation $\tilde{\eta}_6$ (and to a lesser degree also the $5$-loop resummation $\tilde{\eta}_5$) varies only very little.
If we further increase $b$, the curves become more sensitive to $\lambda$ again. However, even in the plot for $b=25$, the value at the near-optimal $\lambda = 2.5$ from figure~\ref{fig:b-plots-of-lambda} remains essentially unchanged.

\begin{figure}
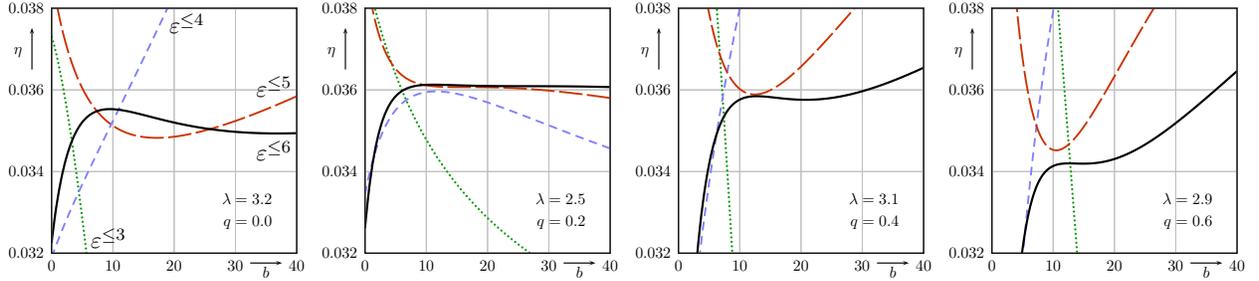

	\centering%
	\includegraphics[width=0.24\textwidth]{{{plots/eta3n1q0lambda3.2}}}~
	\includegraphics[width=0.24\textwidth]{{{plots/eta3n1q0.2lambda2.5}}}~
	\includegraphics[width=0.24\textwidth]{{{plots/eta3n1q0.4lambda3.1}}}~
	\includegraphics[width=0.24\textwidth]{{{plots/eta3n1q0.6lambda2.9}}}
	\caption{For different values of $q \in \set{0,0.2,0.4,0.6}$, we plot the dependence of the resummation on $b$. In each case, we adjusted $\lambda$ to find the best apparent stability with respect to $b$. The loop order $\ell$ is indicated by the label $\eps^{\leq \ell}$.}%
	\label{fig:q-dependence}%
\end{figure}

Finally, it is important to stress the role of the homographic transformation, indexed by $q$. The re-expansion in $\eps'$ after the substitution \eqref{eq:homographic} adjusts the coefficients in a non-trivial way and thereby has a potential to alter the apparent convergence of the resummation procedure. In figure~\ref{fig:q-dependence} we show the dependence on $b$ for various choices of $q$, where in each case $\lambda$ was tuned to minimize the sensitivity to $b$.

The wide plateau existent for $q=0.2$ (already shown in figure~\ref{fig:b-plots-of-lambda}) gets shorter for larger $q$ and the dependence on $b$ becomes much stronger away from a short range of $q$ near $0.2$. Also note that the level of the (shortened) plateaus shifts with $q$ (in figure~\ref{fig:q-dependence}, the plateau moves down when $q$ is raised above $0.2$). Clearly, such a strong correlation of the resummation result with a free parameter is not desirable. But we see that the stability criterion with respect to $b$ nevertheless clearly singles out a preferred range of $q$.
The same qualitative observation applies to the dependence on $\lambda$, that is, for $q$ far from $0.2$, the dependence on $\lambda$ (as in the plots like the ones shown in figure~\ref{fig:lambda-plots-of-b}) becomes stronger.

In summary, we confirm the observation of \cite{GuillouZinnJustin:Accurate} that the homographic transformation \eqref{eq:homographic}, with a suitable choice of $q$, can significantly improve the apparent stability of the resummation with respect to variations in $b$ and $\lambda$. Incidentally, note that the nearly optimal choice (with respect to these stabilities) of setting $(b,\lambda,q)$ to $(15,2.5,0.2)$ yields a result of $\tilde{\eta}^{b,\lambda,q}_6 \approx 0.03611$, which agrees well with earlier resummations and estimates from other methods (see table~\ref{tab:exponents-3d}).
Without the homographic transformation, that is $q=0$, the stability and the agreement with other results would be much worse (see the left-most plot in figure~\ref{fig:q-dependence}).

\subsection{Resummation algorithm}
\label{sec:resummation-algorithm}

In order to quantify the sensitivity of a function $F(x)$ with respect to a resummation parameter $x$, we pick a scale $\Delta_x$ and define
\begin{equation}
	\Variation{x}\left( F(x) \right)
	\defas
		\min_{a\leq x \leq a+\Delta_x} 
		\left(
			\max_{a \leq x' \leq a+\Delta_x}
			\abs{F(x)-F(x')}
		\right)
	\label{eq:variation}%
\end{equation}
as the minimum spread of the values $F(x')$ around $F(x)$ inside an interval $x' \in [a,a+\Delta_x]$ of width $\Delta_x$ that contains $x$ (so $a$ runs from $x-\Delta_x$ to $x$). A smaller value of this quantity corresponds to an increasingly flat plateau (of width $\Delta_x$) in the kind of plots we show above.
Following our discussion in section~\ref{sec:parameter-dependence}, we want to pick the resummation parameters such that these spreads, and in particular $\Variation{b}( \tilde{f}_{\ell}^{b,\lambda,q} )$, become as small as possible.
But this is not the only desirable criterion.

A further indicator for the uncertainty of the resummation is given by the size of the corrections going from one loop order to the next. In fact, in \cite{VladimirovKazakovTarasov:Calculation,GorishnyLarinTkachov:O(eps5)}, $\lambda$ was determined exclusively by minimizing the relative differences
\begin{equation*}
	Q_{\ell} \defas \abs{1-{\tilde{f}_{\ell}}/{\tilde{f}_{\ell-1}}}
\end{equation*}
of the last few loop orders. Pictorially, this amounts to searching for intersections of the curves labelled $\eps^{\leq 6}$, $\eps^{\leq 5}$ and $\eps^{\leq 4}$ in the plots as shown in figures~\ref{fig:b-plots-of-lambda} and \ref{fig:lambda-plots-of-b}.

These two very general approaches are called \emph{principle of minimum sensitivity} (PMS) \cite{Stevenson:ResolutionAmbiguity} and \emph{principle of fastest apparent convergence} (PFAC) \cite{Grunberg:RGimprovedQCD}. In our context of critical exponents, they are discussed for example in \cite{DelamotteDudkaHolovatchMouhanna:Frustrated}, and they can be applied in many different ways. In fact, we found that several works on resummations leave out some of the details of the employed procedure, making it difficult to reproduce the results.\footnote{%
	Some notable exceptions are \cite{CarmonaPelissettoVicari:GinzburgLandau} and \cite[Appendix~A]{ButtiToldin:N>4}, where the scanned parameter ranges are discussed in detail.
}
Therefore, let us state our method precisely.

We combine both, the PMS and the PFAC, into the \emph{error estimate}
\begin{multline}
	E^f_{\ell} (b,\lambda,q)
	\defas
	\max \set{
		\abs{\tilde{f}^{b,\lambda,q}_{\ell}-\tilde{f}^{b,\lambda,q}_{\ell-1}},
		\abs{\tilde{f}^{b,\lambda,q}_{\ell}-\tilde{f}^{b,\lambda,q}_{\ell-2}}
	}
	\\
	+ \max \set{
		\Variation{b} \left( \tilde{f}^{b,\lambda,q}_{\ell} \right)
		,\Variation{b} \left( \tilde{f}^{b,\lambda,q}_{\ell-1} \right)
	  }
	  + \Variation{\lambda} \left( \tilde{f}^{b,\lambda,q}_{\ell} \right)
	  + \Variation{q} \left( \tilde{f}^{b,\lambda,q}_{\ell} \right)
	.
	\label{eq:error-estimate}%
\end{multline}
The first term accounts for the uncertainty due to the unknown corrections from higher perturbative orders (estimated by the differences of the largest two loop orders), whereas the spreads in the second line take care of the arbitrariness in the choice of the parameters $b$, $\lambda$ and $q$.
We pick the scales as follows:
\begin{itemize}
	\item $\Delta_b = 20$, because we indeed find such strikingly wide plateaus (with only minute variations of the resummation result), as seen in figure~\ref{fig:b-plots-of-lambda} and noticed in \cite{MudrovVarnashev:ModifiedBorel}, for all critical exponents and values of $\n$ that we considered.

	\item $\Delta_{\lambda} = 1$, since the dependence on $\lambda$ is stronger and even at six loops the plateaus do not grow much longer (see figure~\ref{fig:lambda-plots-of-b}).

	\item $\Delta_{q} = 0.02$, as the dependence on $q$ is very strong (for fixed $\lambda$); higher values of $\Delta_q$ would yield unrealistically high error estimates.
\end{itemize}
Our resummation of a critical exponent $f$ at order $\ell$ then proceeds as follows:
\begin{enumerate}
	\item Sample $\tilde{f}^{b,\lambda,q}_{\ell}$ as defined in \eqref{eq:borel-inverse-homographic} for parameters in the cube 
		\begin{equation*}
			(b,\lambda,q) \in [0,40] \times [0,4.5] \times [0,0.8],
		\end{equation*}
		where $b$ runs over half-integers and $\lambda$ and $q$ are probed in steps of $0.02$.

	\item For each such point $(b,\lambda,q)$, compute the error estimate \eqref{eq:error-estimate}.
		
	\item Find the (global) minimum $\overline{E}^f_{\ell}$ of these values for $E^f_{\ell}(b,\lambda,q)$.
\end{enumerate}

For the example of $\eta$ (in the $3$-dimensional Ising case $\n=1$), the ``apparently best'' resummation, according to our error estimate, occurs at $(b,\lambda,q)=(11,2.56,0.2)$ and yields $\overline{E}^{\eta}_{6} \approx 0.0002$. These resummation parameters are very close to the second plots in figures~\ref{fig:b-plots-of-lambda} and \ref{fig:lambda-plots-of-b}. The actual value of the resummed critical exponent is
$\tilde{\eta}_6^{11,2.56,0.2} \approx 0.03615$
and agrees (within the error estimate $\overline{E}^{\eta}_6$) with results from completely different theoretical approaches (see table~\ref{tab:exponents-3d}).

A comment is due on our choice for the function $E^f_{\ell}(b,\lambda,q)$, which we use as a quantitative measure for the ``quality'' of a resummation. Obviously, there are many different reasonable definitions of such a measure.\footnote{%
	For example, one could incorporate more low order corrections $|\tilde{f}_{\ell}-\tilde{f}_{\ell-k}|$ with $k\geq 2$, or disregard the stability $\Variation{b} (\tilde{f}_{\ell-1})$, or take further stabilities of lower loop orders into account.
}
And even if we stay with our definition~\eqref{eq:error-estimate}, it still depends on the somewhat arbitrary parameters $\Delta_b$, $\Delta_{\lambda}$ and $\Delta_q$.
However, we tested numerous such variations and found that their effect on the selection of the ``apparently best'' resummation parameters $(b,\lambda,q)$ only results in very small shifts of the critical exponents.
David Broadhurst attributes a fitting quote to Jean Zinn-Justin (paraphrased):
\begin{quote}
In work on resummation, there is always an undeclared parameter: the number of methods tried and rejected before the paper was written.
\end{quote}
We stopped counting.

\subsection{Error estimates}
\label{sec:error-estimates}

\begin{figure}
	\centering
	\includegraphics[height=5cm]{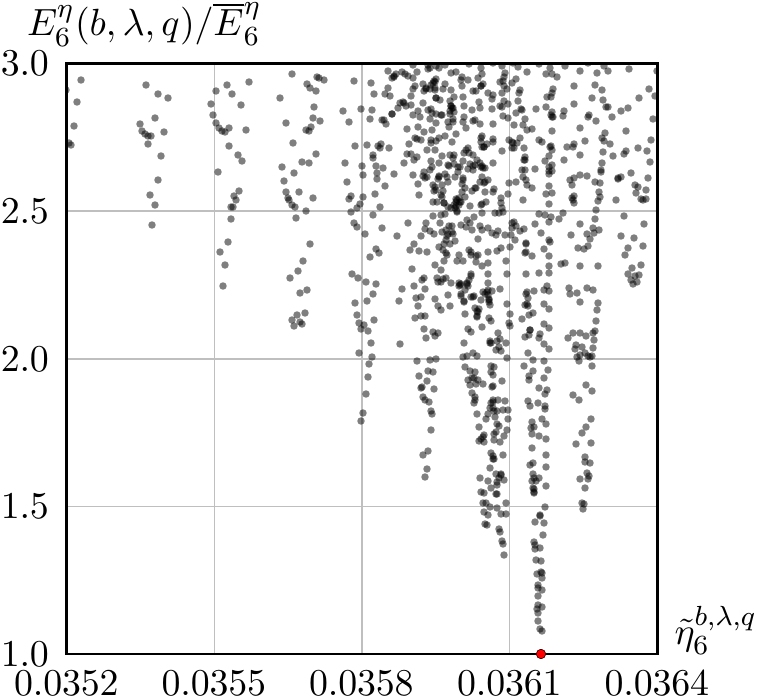}\qquad
	\includegraphics[height=5cm]{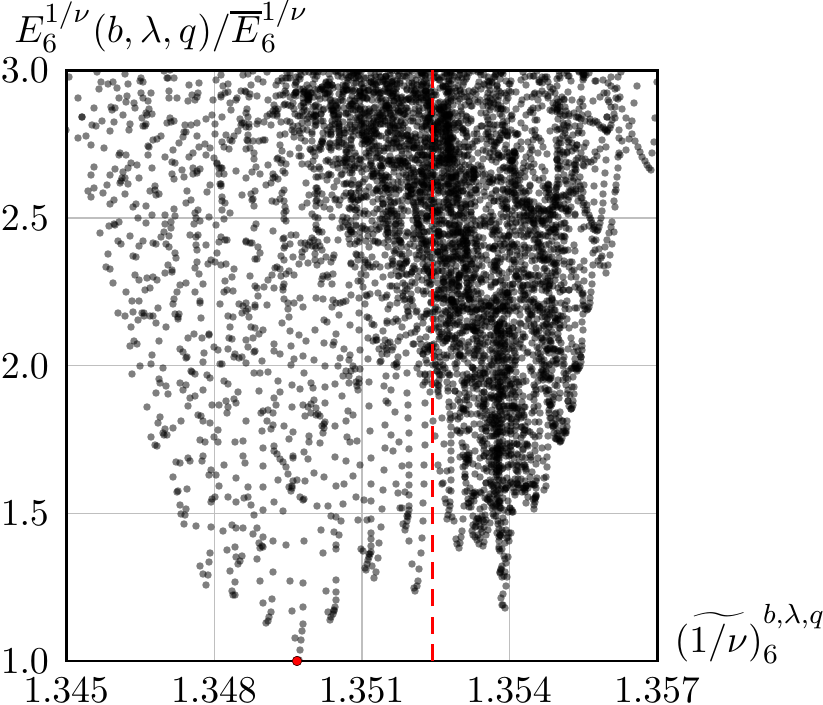}
	\caption{
		Six loop resummation results and their error estimates \eqref{eq:error-estimate} for the critical exponent $\eta$ of the Ising model (left plot) and $\nu^{-1}$ of the $O(4)$ model (right plot), in $3$ dimensions.
		In the first case, the ``best'' resummation is rather sharply localized, whereas the second plot reveals a wide spread among the resummations with small error estimates. The dashed vertical line shows the weighted average.
	}%
	\label{fig:resummation-spreads}%
\end{figure}

The estimation of resummation errors is a notoriously difficult undertaking, for mainly two reasons:
\begin{itemize}
	\item We do not know the perturbative coefficients of the next loop order.
	\item The free parameters (in our case: $b$, $\lambda$ and $q$) can in principle be tuned to reproduce any value for the critical exponents.
\end{itemize}
Therefore, we can only hope to get a good guess of the error if we assume:
\begin{enumerate}
	\item[(A)] The next perturbative correction is not much larger than the last known correction.
	\item[(B)] The exact critical exponent is close to the resummed critical exponent $\tilde{f}_{\ell}^{b,\lambda,q}$ when the parameters $(b,\lambda,q)$ are chosen in order to minimize $E^f_{\ell}(b,\lambda,q)$.
\end{enumerate}
We chose the widths $\Delta_b$, $\Delta_{\lambda}$ and $\Delta_q$ as given above such that $E^f_{\ell}(b,\lambda,q)$ from \eqref{eq:error-estimate} should be considered as a lower bound on the error that is inherent to the resummation $\tilde{f}_{\ell}^{b,\lambda,q}$. To verify that this guess of the error is self-consistent, we consider plots as shown in figure~\ref{fig:resummation-spreads}, i.e.\ the critical exponent $\eta$ of the $3$-dimensional Ising model ($\n=1$).
Each point $(\tilde{\eta}_6^{\theta}, E^{\eta}_6(\theta))$ represents a set $\theta=(b,\lambda,q)$ of resummation parameters with an error estimate $\leq 0.0006 \approx 3 \overline{E}^{\eta}_6$. We notice that the optimal resummation ($\tilde{E}^{\eta}_6 \approx 0.0002$) is rather sharply localized. The spread of the resummation results around this optimum increases in line with the error estimate---this means that our error estimates are indeed consistent with the actual spread.

But there are also cases, like the critical exponent $\nu$ in the $O(4)$ model as illustrated in the right plot of figure~\ref{fig:resummation-spreads}, where there exist many resummation parameters that yield a close to minimal error estimate, but which produce results that are spread much more widely than the error estimate suggests. In such a situation, we do not pick the ``apparently best'' resummation (like we do for $\eta$), but instead we choose a mean value (e.g.\  $1.351$ in the $O(4)$ example) as a more faithful representation of the distribution of resummations. This is how we compiled our resummation data in table~\ref{tab:exponents-3d} in the next section. Similar plots are provided for all exponents in an ancillary file (see appendix~\ref{sec:files}).

In order to take the actual spreads of the best resummations (like shown in figure~\ref{fig:resummation-spreads}) into account, all errors reported in the following section consist of the minimum error estimate $\overline{E}^f_{\ell}$, plus an additional contribution given by two standard deviations of the set of all resummations with an error $\leq 3\overline{E}^f_{\ell}$. 
One might argue that this would overestimate the error, since we aimed to account for the spreads due to variations in the parameters $(b,\lambda,q)$ already in \eqref{eq:error-estimate}.
However, given the arbitrariness in fixing the widths $\Delta_x$ and the unknown uncertainties neglected by assuming (A) and (B) above, we are more confident with stating these enlarged errors (in particular since underestimated errors are not uncommon in the literature \cite[last paragraph]{JaschKleinert:Fast3}).


\begin{figure}
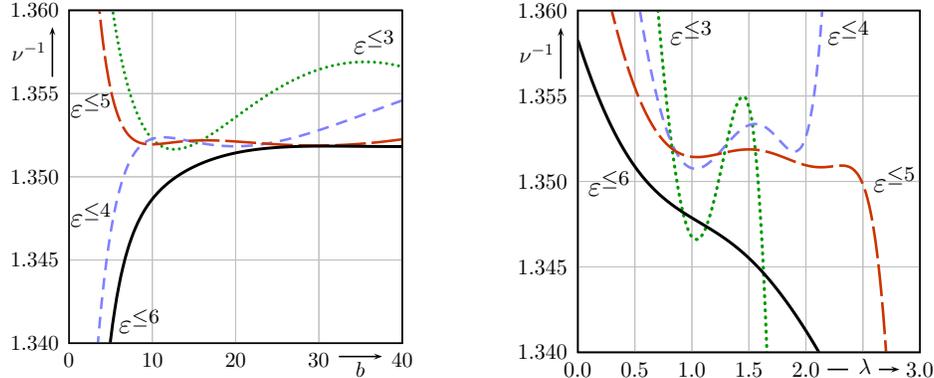

	\centering
	\includegraphics[height=5cm]{{{plots/invnu3n4q0.4lambda0.82}}}\qquad\qquad
	\includegraphics[height=5cm]{{{plots/invnu3n4q0.4b10}}}
	\caption{Dependence of the resummation results for the critical exponent $\nu^{-1}$ in the $3$-dimensional $O(4)$ model on $b$ and $\lambda$, around the point $(b,\lambda,q)=(10,0.82,0.4)$ of the ``apparently best'' $5$-loop resummation. In this case, the $6$th loop order induces a significant correction, which exceeds what one might expect from just considering the variation (as a function of $b$ and $\lambda$) intrinsic to the $5$-loop resummation itself.}%
	\label{fig:wide-deceptive-convergence}%
\end{figure}

Let us point out in an explicit example how the error predicted by the principles PMS and PFAC, like \eqref{eq:error-estimate}, can underestimate the corrections from higher order contributions.
We consider again the exponent $\nu$ in the 3-dimensional $O(4)$ model.
The apparently best $5$-loop resummation gives $\nu^{-1} \approx 1.352$ at $\theta=(b,\lambda,q)=(10, 0.82, 0.4)$ with an error estimate of $\overline{E}^{1/\nu}_5 \approx 0.002$.
On the scale of figure~\ref{fig:wide-deceptive-convergence}, we see indeed only very small fluctuations of the $5$-loop resummation when $b$ and $\lambda$ are varied. Furthermore, the $3$-, $4$- and $5$-loop resummations essentially coincide at $\theta$; in other words, the $4$- and $5$-loop corrections almost vanish at this point.
Nonetheless, we see that the $6$-loop correction is significant and much larger than the fluctuations of the $5$-loop result. This kind of behaviour appears to be linked with large spreads of the best resummations as shown in the right plot of figure~\ref{fig:resummation-spreads}, and explains why, in some rare cases, our error estimates for the $6$-loop resummations quoted in table~\ref{tab:exponents-3d} actually exceed the error estimates for the resummations of the 5-loop series.

In such a case one could say that the $5$-loop result \emph{seemed converged}, but towards an erroneous value. This phenomenon is particularly well-known to occur in two dimensions, where it was called ``anomalous apparent convergence'' in \cite{DelamotteDudkaHolovatchMouhanna:Frustrated}.
Here we also like to stress that the large $x$ behaviour of the Borel transform $\Borel[b,\lambda,\ell]{f}(x)$ of critical exponents $f$ is still unknown. Hence the power law model \eqref{eq:borel-lambda-ansatz} is only justified because it seems to work very well in practice, but it remains unclear if $\lambda$ can actually be interpreted as the exponent of a power law asymptotic behaviour of $f(\eps)$ for large $\eps$. If the exact large $x$ behaviour is of a different form, then the PMS might miss the correct value \cite[Appendix~A]{Kleinert:ConvertingWeakStrong}.


\section{Estimates for critical exponents in $3$ dimensions}
\label{sec:3dim-results}

The critical phenomena of many interesting physical systems are described by the $O(\n)$ universality classes. We refer to \cite{PelissettoVicari:CriticalPhenomena} for a comprehensive discussion, and only recall some of the most interesting examples:
\begin{itemize}
	\item[$\n=0$] (self-avoiding walks): polymers \cite{Gennes:ScalingPolymers},

	\item[$\n=1$] (Ising universality class): liquid-vapour transitions, uniaxial magnets,

	\item[$\n=2$] (XY universality class): superfluid $\lambda$-transition of helium \cite{LipaNissenStrickerSwansonChui:SpecificHeat},

	\item[$\n=3$] (Heisenberg universality class): isotropic ferromagnets,

	\item[$\n=4$\rlap{:}] \ finite temperature QCD with two light flavours \cite{PisarskiWilcez:ChiralChromo}.
\end{itemize}
These are the systems that we consider below; larger values of $\n$ were discussed for example in \cite{ButtiToldin:N>4,AntonenkoSokolov:O(n)>3,HoltmannSchulze:O(6)}.

\begin{table}
	\caption{Estimates for critical exponents in $\D=3$ dimensions of the $O(\n)$ vector model. Results from the conformal bootstrap and Monte Carlo techniques are listed first (we tried to collect the most accurate predictions in each case). Our estimates from the $5$- and $6$-loop $\eps$-expansions are shown next. For comparison of the resummation methods, we display the $5$-loop results (from $\eps$-expansion without $\D=2$ boundary conditions) according to \cite{GuidaZinnJustin:CriticalON}.}%
	\label{tab:exponents-3d}%
	\centering%
	\begin{minipage}{\textwidth}\renewcommand{\footnoterule}{\vspace{-2mm}}
	\centering%
	\begin{tabular}{@{}r@{}rl@{ }l@{ }l@{ }l@{ }l}
	\toprule
&   & {$\n=0$} & {$\n=1$} & {$\n=2$} & {$\n=3$} & {$\n=4$} \\
\midrule
\llap{\ldelim\{{4}{*}[{$\eta$}]} &    
    & 0.031043(3)\footnote{From $\gamma=1.156953(1)$ \cite{Clisby:ScaleFreeSAW} and $\nu=0.5875970(4)$ \cite{ClisbyDuenweg:Hydrodynamic} via $\gamma=\nu(2-\eta)$ in \eqref{eq:scaling-relations}.} 
    & 0.036298(2)\textsuperscript{\cite{KosPolandDuffinVichi:Islands}} 
    & 0.0381(2)\textsuperscript{\cite{CampostriniHasenbuschPelissettoVicari:4He}} 
    & 0.0378(3)\textsuperscript{\cite{HasenbuschVicari:Anisotropic}} 
    & 0.0360(3)\footnote{Given in \cite{HasenbuschVicari:Anisotropic} and compatible with $0.0365(10)$ \cite{Hasenbusch:EliminatingN34} and $y_h=(5-\eta)/2=2.4820(2)$ in \cite{Deng:O(4)}.}
\\
&  $\eps^6$ 
    & 0.0310(7)  & 0.0362(6)  & 0.0380(6)  & 0.0378(5)  & 0.0366(4)  \\
&  $\eps^5$ 
    & 0.0314(11) & 0.0366(11) & 0.0384(10) & 0.0382(10) & 0.0370(9)  \\
&  {\cite{GuidaZinnJustin:CriticalON}}
    & 0.0300(50) & 0.0360(50) & 0.0380(50) & 0.0375(45) & 0.036(4)  \\
%
\cline{2-7}
\llap{\ldelim\{{4}{*}[{$\nu$}]} &    
    & 0.5875970(4)\textsuperscript{\cite{ClisbyDuenweg:Hydrodynamic}} 
    & 0.629971(4)\textsuperscript{\cite{KosPolandDuffinVichi:Islands}}
    & 0.6717(1)\textsuperscript{\cite{CampostriniHasenbuschPelissettoVicari:4He}}
    & 0.7112(5)\textsuperscript{\cite{CampostriniHasenbuschPelissettoRossiVicari:Heisenberg}}
    & 0.7477(8)\footnote{From $y_t=1/\nu=1.3375(15)$ in \cite{Deng:O(4)}, compatible with $\nu=0.749(2)$ \cite{Hasenbusch:EliminatingN34} and $0.750(2)$ \cite{HasenbuschVicari:Anisotropic}.}
\\
&   $\eps^6$ 
   & 0.5874(3)  & 0.6292(5)  & 0.6690(10) & 0.7059(20) & 0.7397(35) \\
&   $\eps^5$ 
   & 0.5873(13) & 0.6290(20) & 0.6687(13) & 0.7056(16) & 0.7389(24) \\
&   {\cite{GuidaZinnJustin:CriticalON}}
   & 0.5875(25) & 0.6290(25) & 0.6680(35) & 0.7045(55) & 0.737(8)   \\
%
\cline{2-7}
\llap{\ldelim\{{4}{*}[{$\omega$}]} &    
    & 0.904(5)\footnote{Computed from $\omega\nu=\Delta=0.531(3)$ according to \cite{Belohorec:PhD} 
    and $\nu=0.5875970(4)$ in \cite{ClisbyDuenweg:Hydrodynamic}.}
    & 
	    0.830(2)\textsuperscript{\cite{3dIsingBootstrapII}}
    & 
	    0.811(10)\extrafootmark{e}
    & 0.791(22)\extrafootmark{e} 
    & 0.817(30)\footnote{These are the results given as $\Delta_{S'}=3+\omega$ in \cite[Table~2]{EcheverriHarlingSerone:EffectiveBootstrap}.}%
\\
&  $\eps^6$ 
    & 0.841(13) & 0.820(7)  & 0.804(3)  & 0.795(7)  & 0.794(9) \\
&  $\eps^5$ 
    & 0.835(11)  & 0.818(8)  & 0.803(6)  & 0.797(7)  & 0.795(6) \\
&  {\cite{GuidaZinnJustin:CriticalON}}
    & 0.828(23) & 0.814(18) & 0.802(18) & 0.794(18) & 0.795(30) \\
\bottomrule
	\end{tabular}
\end{minipage}
\end{table}

The critical behaviour of each universality class is governed by the critical exponents $\alpha$, $\beta$, $\gamma$, $\delta$, $\eta$ and $\nu$. However, in our field theoretic approach, only two of them are independent and determine all others through the scaling relations \eqref{eq:scaling-relations}. So, while discussing agreement with other theoretical methods, we will only consider the critical exponents $\eta$, $\nu$ and the correction to scaling exponent $\omega$.

In table~\ref{tab:exponents-3d}, we present the summary of our results for the six loop resummation of these exponents, in $\D=3$ dimensions for $0\leq\n\leq 4$.
For comparison of the resummation methods, 
we also show the outcome of applying our resummation procedure to the five-loop $\eps$-expansions, in comparison with the results given in \cite{GuidaZinnJustin:CriticalON} for the summation of the same series.
It should be noted that we do not consider the renormalization in fixed dimension $\D\in\set{2,3}$, where the resummation problem is slightly different and has been approached, for example, with the \emph{pseudo-$\eps$ expansion} \cite{NikitinaSokolov:Pseudo2,SokolovNikitina:FisherPseudo} going back to Nickel \cite[reference~19]{LeGuillouZinnJustin:CriticalFromField}.

The table furthermore includes a tiny selection of estimates obtained with other theoretical approaches (in particular Monte Carlo and the conformal bootstrap), which by no means can represent the vast literature on this subject. We merely tried to pick the most recent results of the highest apparent accuracy, in order to compare them against our field theoretic method.

Let us first state the following general observations:
\begin{enumerate}
	\item When we apply our resummation algorithm to the five loop $\eps$-expansions, we obtain values for the critical exponents that are compatible with the resummation of \cite{GuidaZinnJustin:CriticalON}. 
	This indicates that our resummation procedure is consistent with their method, though it differs from ours.\footnote{%
		Our method from section~\ref{sec:resummation} is an implementation of the ideas lined out in \cite{GuillouZinnJustin:Accurate}.
		In \cite{GuidaZinnJustin:CriticalON}, however, the authors abandoned the parameter $\lambda$ from \eqref{eq:borel-lambda-ansatz} and in its stead introduced a further parameter $r$ via a transformation $f(\eps)\mapsto (1+r\eps)f(\eps)$.
	}

	\item Our error estimates (at 5 loops) are smaller than those given in \cite{GuidaZinnJustin:CriticalON}.

	\item The $6$-loop resummation results are consistent with the $5$-loop results (within the quoted errors), and in most cases the apparent errors of the $6$-loop results are significantly reduced compared to $5$ loops.

	\item Overall the agreement of the $6$-loop resummation with predictions from other theoretical approaches is very good, in particular for $\eta$.
\end{enumerate}
The largest apparent discrepancies between our $6$-loop resummation and other estimates occur for $\omega$ at $\n=0$ and the exponent $\nu$ when $\n \geq 1$.
In fact, the trend that renormalization group based predictions for $\nu$ tend to be lower than results from statistical approaches has been observed long ago. Our six loop results narrow this gap only slightly, but due to the large error estimates in those cases we do not attach any significance to these differences yet. Once the $7$ loop perturbative results become available, it will be interesting to revisit these cases.

We will now briefly discuss the universality classes one-by-one and, for completeness, show the full set of critical exponents as obtained via the scaling relations \eqref{eq:scaling-relations} from our resummation results for $\eta$ and $\nu$. However, these derived exponents might be determined more accurately via direct resummations of the individual series, as in \cite{GuidaZinnJustin:CriticalON}, or other resummation techniques.

Note that we resum the $\eps$-expansions as explained in section~\ref{sec:resummation}, without enforcing any boundary values of exactly known critical exponents in two dimensions. The latter technique is often used to improve the resummation results for three dimensions \cite{GuidaZinnJustin:CriticalON,GuillouZinnJustin:Accurate,Gracey:4loopPhi3}. However, the exact boundary values are not known in all cases, and it seems difficult to quantify the effect of this procedure on the error estimates.
We therefore do not enforce any two-dimensional boundary values; instead, we test our method in section~\ref{sec:2dim-results} by comparing our resummation results in two dimensions with exact predictions.

\subsection{Self-avoiding walks ($\n=0$)}

Over the last decade, successive improvements of Monte Carlo methods significantly diminished the uncertainty of critical exponents \cite{ClisbyLiangSlade:SAWlace,Clisby:SAWfastpivot,Clisby:ScaleFreeSAW,ClisbyDuenweg:Hydrodynamic}.
The latest and most accurate estimates are $\gamma=1.156953(1)$ \cite{Clisby:ScaleFreeSAW} and $\nu=0.5875970(4)$ \cite{ClisbyDuenweg:Hydrodynamic}. In contrast, the value $\omega\nu=\Delta=0.531(3)$, computed long ago in \cite{Belohorec:PhD} and confirmed by the very recent result $0.528(8)$ of \cite{ClisbyDuenweg:Hydrodynamic}, remains the most precise determination of the correction to scaling.
In conclusion, we derive
\begin{equation}
	\eta = 2-\frac{\gamma}{\nu} = 0.031043(3)
	\quad\text{and}\quad
	\omega = \frac{\Delta}{\nu} = 0.904(5).
	\label{eq:n=0-MC}%
\end{equation}
Applying the resummation procedure described in section~\ref{sec:resummation} to the six loop $\eps$-expansions of $\eta$, $\nu$ and $\omega$ (tables \ref{tab:eta-numeric}--\ref{tab:omega-numeric}) yields
\begin{equation}
	\eta = 0.0310(7), \quad \nu = 0.5874(3) \quad\text{and}\quad \omega = 0.841(13).
	\label{eq:n=0-eps}%
\end{equation}
The values of $\eta$ and $\nu$ are in good agreement with \eqref{eq:n=0-MC}, but the correction to scaling exponent $\omega$ differs by $\approx 7\%$. Note that our error estimate for $\omega$ increases from $5$ to $6$ loops, which hints towards a badly convergent situation.

For completeness, we compute the other critical exponents via the scaling relations \eqref{eq:scaling-relations} from \eqref{eq:n=0-eps}:
\begin{equation}
	\alpha = 0.2378(9),\quad
	\beta  = 0.3028(4),\quad
	\gamma = 1.1566(10),\quad
	\delta = 4.820(4).
\end{equation}

\subsection{Ising universality class ($n=1$)}

Experimental measurements in Ising systems, as discussed for example in \cite{SengersShanks:Fluids,LytleJacobs:Turbidity,PelissettoVicari:CriticalPhenomena}, have rather larger uncertainties.
Theoretical predictions are more accurate, like the Monte Carlo simulations \cite{Hasenbusch:ScalingLattice3dIsing} with
\begin{equation}
	\eta   = 0.03627(10),\quad
	\nu    = 0.63002(10),\quad 
	\omega = 0.832(6).
\end{equation}
The most accurate values were obtained with the conformal bootstrap \cite{KosPolandDuffinVichi:Islands,3dIsingBootstrapII}:
\begin{equation}
	\eta   = 0.036298(2),\quad
	\nu    = 0.629971(4), \quad 
	\omega = 0.830(2).
\end{equation}
Our resummations for $\eta$, $\nu$ and $\omega$ and the other exponents derived via \eqref{eq:scaling-relations} are
\begin{equation}\begin{split}
	\eta   = 0.0362(6), \quad
	\nu    = 0.6292(5), \quad 
	\omega = 0.820(7) \quad\text{and}
	\\
	\alpha = 0.112(2),\quad
	\beta  = 0.3260(5),\quad
	\gamma = 1.2356(14),\quad
	\delta = 4.790(4).
\end{split}\end{equation}

\subsection{XY universality class ($n=2$)}

Famous for the very precise measurement $\alpha= -0.0127(3)$ in the microgravity experiment at the $\lambda$-transition of liquid helium~\cite{LipaNissenStrickerSwansonChui:SpecificHeat}, this universality class also describes planar Heisenberg magnets.
Theoretical predictions from a combination of Monte Carlo simulations and High-Temperature expansions in \cite{CampostriniHasenbuschPelissettoVicari:4He} are
\begin{equation}
\begin{split}
	\eta   =  0.0381(2), \quad
	\nu    =  0.6717(1), \quad 
	\omega =  0.785(20)
	\\
	\alpha = -0.0151(3), \quad 
	\beta  =  0.3486(1), \quad 
	\gamma =  1.3178(2), \quad
	\delta =  4.780(1).
\end{split}
\end{equation}
The conformal bootstrap \cite{EcheverriHarlingSerone:EffectiveBootstrap} provides a correction to scaling exponent $\omega=\Delta_{S'}-3=0.811(10)$.
Resumming the six loop $\eps$-expansions, we obtain
\begin{equation}\begin{split}
	\eta   = 0.0380(6),\quad
	\nu    = 0.6690(10),\quad
	\omega = 0.804(3),
	\\
	\alpha =-0.007(3),\quad
	\beta  = 0.3472(7),\quad
	\gamma = 1.313(2),\quad 
	\delta = 4.780(3),
\end{split}\end{equation}
where the values in the second row are calculated with the scaling relations \eqref{eq:scaling-relations}. The accuracy for $\alpha=2-3\nu$ is so small due to the vicinity of $\nu$ and $2/3$, and serves another motivation for the seven loop calculation of the $\eps$-expansion.

\subsection{Heisenberg universality class ($n=3$)}
	For experimental results, we refer to \cite{GRRNEBDM:Nd}. 
	The most precise theoretical predictions stem from Monte Carlo simulations~\cite{HasenbuschVicari:Anisotropic}:
\begin{equation}
	\eta   = 0.0378(3),\quad
	\nu    = 0.7116(10);
\end{equation} 
Monte Carlo combined with High-Temperature expansion~\cite{CampostriniHasenbuschPelissettoRossiVicari:Heisenberg}:
\begin{equation}
\begin{split}
	\eta   = 0.0375(5),\quad
	\nu    = 0.7112(5),
	\\
	\alpha =-0.1336(15),\quad
	\beta  = 0.3689(3),\quad
	\gamma = 1.3960(9),\quad
	\delta = 4.783(3);
\end{split}
\end{equation}
and the correction to scaling exponent $\omega=\Delta_{S'}-3=0.791(22)$ from the conformal bootstrap \cite{EcheverriHarlingSerone:EffectiveBootstrap}.
Our resummations yield
\begin{equation}\begin{split}
	\eta   = 0.0378(5),\quad
	\nu    = 0.7059(20),\quad
	\omega = 0.795(7),
	\\
	\alpha =-0.118(6),\quad
	\beta  = 0.3663(12),\quad
	\gamma = 1.385(4),\quad
	\delta = 4.781(3).
\end{split}\end{equation}

\subsection{The case $n=4$}
The Monte Carlo results
$\eta = 0.0360(3)$, $\nu = 0.750(2) $
from \cite{HasenbuschVicari:Anisotropic} and
$ \eta = 0.0365(10)$, $\nu = 0.749(2)$
given in \cite{Hasenbusch:EliminatingN34} are consistent with each other and also with the values
$\eta = 0.0360(4)$, $\nu  = 0.7477(8)$
obtained via
$ y_t=1/\nu$ and $y_h=(5-\eta)/2$ from the results in \cite{Deng:O(4)}.
The correction to scaling exponent is $\omega=\Delta_{S'}-3=0.817(30)$ according to the conformal bootstrap \cite{EcheverriHarlingSerone:EffectiveBootstrap}.

Our resummation results and the scaling relations \eqref{eq:scaling-relations} lead to
\begin{equation}\begin{split}
	\eta   = 0.0366(4),\quad
	\nu    = 0.7397(35),\quad
	\omega = 0.794(9) \quad\text{and}
	\\
	\alpha = -0.219(11),\quad
	\beta  =  0.383(2),\quad
	\gamma =  1.452(7),\quad
	\delta =  4.788(2).
\end{split}\end{equation}

\section{Critical exponents in two dimensions}
\label{sec:2dim-results}

\begin{table}
\caption{%
	Previous estimates for $\D=2$ dimensions from the $5$-loop $\eps$-expansion according to \cite{GuillouZinnJustin:Accurate}, our results for the resummation of the same series and also for six loops.
	The first row for the exponents $\eta$ and $\nu$ shows the exact values for the Ising model (column $\n=1$) due to Onsager \cite{Onsager:CrystalStatisticsI} and the conjectures of Nienhuis \cite{Nienhuis:ExactO(n)} in the cases $\n=-1$ and $\n=0$.
	We also report theoretical expectations for the correction to scaling exponent $\omega$.
}%
	\label{tab:exponents-2d}%
	\centering
	\begin{tabular}{@{}r@{}rlll}
\toprule
       &   & $\n=-1$ & $\n=0$ & $\n=1$  \\
\midrule
\llap{\ldelim\{{4}{*}[{$\eta$}]}    
	& \cite{Nienhuis:ExactO(n)} & $0.15$ 
				    & $0.208333\ldots$ 
				    & $0.25$ \\
	& $\eps^6$ &  0.130(17) & 0.201(25) & 0.237(27)   \\
	& $\eps^5$ &  0.137(23) & 0.215(35) & 0.249(38)   \\
 	& \cite{GuillouZinnJustin:Accurate} & & 0.21(5) & 0.26(5) \\
\cline{2-5}
\llap{\ldelim\{{4}{*}[{$\nu$}]}    
	& \cite{Nienhuis:ExactO(n)} &  $0.625$ 
				    &  $0.75$  
				    &  $1$      \\
	& $\eps^6$ & 0.6036(23) & 0.741(4)  & 0.952(14) \\
	& $\eps^5$ & 0.6025(27) & 0.747(20) & 0.944(48) \\
	&\cite{GuillouZinnJustin:Accurate} & & 0.76(3) & 0.99(4) \\
\cline{2-5}
\llap{\ldelim\{{4}{*}[{$\omega$}]}   
	&&
		& 2\textsuperscript{\cite{Nienhuis:ExactO(n),CGJPRS:Scaling2dSAW}} 
		& 1.75\textsuperscript{\cite{CalabreseCaselleCeliPelissettoVicari:NonAnalyticity}} \\
	& $\eps^6$ &  1.95(28) & 1.90(25) & 1.71(9)  \\
	& $\eps^5$ &  1.88(30) & 1.83(25) & 1.66(11) \\
	&\cite{GuillouZinnJustin:Accurate} & & 1.7(2)  & 1.6(2) \\
\bottomrule
	\end{tabular}
\end{table}

In two dimensions, the resummation of critical exponents is known to be much less accurate, most likely due to non-analyticities in the beta function at the critical point \cite{CalabreseCaselleCeliPelissettoVicari:NonAnalyticity}.\footnote{
	It seems, however, that the $\eps$-expansion yields much more accurate predictions for critical exponents in two dimensions (see table~\ref{tab:exponents-2d}) than the fixed dimension approach \cite{OrlovSokolov:Critical2dim5loop}.
}
Indeed, our errors (determined automatically by the procedure from section~\ref{sec:error-estimates}) reflect this expectation.

The results shown in table~\ref{tab:exponents-2d} are again compatible with the $5$-loop resummations from \cite{GuillouZinnJustin:Accurate}.
Furthermore, we can compare them with the following predictions:
\begin{itemize}
	\item[$\n=1$] The exact critical exponents $\eta=1/4$ and $\nu=1$ of the Ising model were computed by Onsager \cite{Onsager:CrystalStatisticsI}. The convergence of our perturbative results seems slow, and in particular $\nu$ seems to stabilize at a value slightly lower than expected.
This phenomenon of ``anomalous apparent convergence'' was already discussed in detail in \cite{DelamotteDudkaHolovatchMouhanna:Frustrated} at the $5$-loop level, and seems to persist at $6$ loops.

	The correction to scaling, however, is in good agreement with the prediction $\omega=1.75$ from \cite[equation~(21)]{CalabreseCaselleCeliPelissettoVicari:NonAnalyticity}.

	\item[$\n=0$] Our results are compatible with $\eta=5/24$, $\nu=3/4$ and $\omega=2$ as already conjectured by Nienhuis \cite{Nienhuis:ExactO(n)}, though $\nu$ again seems slightly too small and the uncertainty of $\omega$ is large.
	The value of $\omega$ has been subject to extensive debate \cite{CGJPRS:Scaling2dSAW}, so increased accuracy from the $7$-loop $\eps$-expansion would be particularly desirable here.

	\item[$\n=-1$] The value of $\eta$ is roughly consistent with Nienhuis' $\eta=3/20$, but for $\nu$ the prediction of $5/8$ is very far from our result (almost 10 times our error estimate).

\end{itemize}

\section{Summary and Outlook}
\label{sec:outlook}

After many years of work, new mathematical insights into the structure of Feynman integrals have matured into practically applicable techniques that overcome limitations of traditional approaches so far as to enable progress with the perturbative computation of renormalization group functions.
Finally, after 25 years, we were thus able to improve on the five-loop results \cite{KNFCL:5loopPhi4} of the $O(\n)$-symmetric $\field^4$ model. We like to point out that the primitive graphs relevant to this computation have been essentially known for 30 years \cite{Broadhurst:5loopsbeyond}. So the challenge was not in unknown transcendental numbers beyond zeta values, but the complexity introduced through subdivergences.

Our approach rests on a significantly improved understanding of the parametric representation of Feynman integrals \cite{Brown:PeriodsFeynmanIntegrals,Brown:TwoPoint}, symbolic integration algorithms based on hyperlogarithms \cite{Panzer:HyperIntAlgorithms} and the Hopf algebra \cite{CK:RH1} of renormalization underlying the BPHZ scheme \cite{BrownKreimer:AnglesScales}.
But also other techniques, like graphical functions and single-valued integration \cite{Schnetz:GraphicalFunctions,GolzPanzerSchnetz:GfParam}, can be used for this kind of calculations, as demonstrated by the impressive, independent work of Oliver Schnetz \cite{Schnetz:NumbersAndFunctions}. In fact, these methods are so powerful that even the $7$ loop computation seems now not only feasible but is already underway \cite{Schnetz:NumbersAndFunctions}.
Note that the contributions from $7$-loop graphs without subdivergences were investigated numerically long ago \cite{BroadhurstKreimer:KnotsNumbers,Schnetz:Census} and are by now already known exactly \cite{PanzerSchnetz:Phi4Coaction}.

We are optimistic that these tools will also have further applications. They should be particularly amenable to $\field^3$ theories, whose renormalization only very recently reached the 4-loop level \cite{Gracey:4loopPhi3,Gracey:F4at4loops}.
Fermions and gauge fields provide additional technical difficulties, but even in this very challenging domain, significant progress was achieved recently. Let us just mention the Gross-Neveu model \cite{GraceyLutheSchroeder:4loopGN} and the particularly impressive 5-loop renormalization of QCD \cite{BaikovChetyrkinKuehn:5loopQCD,LutheMaierMarquardSchroeder:CompleteQCD5,RuijlUedaVermaserenVogt:4loopQCDvanishing} and generalizations \cite{BaikovChetyrkinKuehn:5loopFermionGeneral,LutheMaierMarquardSchroeder:5loopGeneralDims,HerzogRuijlUedaVermaserenVogt:YMFermions}. Those computations drew on yet another set of recently improved methods, e.g.\ \cite{Baikov:BaikovMethod,LutheSchroeder:5loopTadpoles,HerzogRuijl:Rstar,RUV:ForcerLL2016}.
 
It is our hope that explicit longer perturbation series, like the ones presented here, will lead to an improved understanding of how they approach their asymptotic behaviour, and provide a sufficiently robust and precise testing ground to evaluate and compare the myriad of resummation methods that have been proposed over time, which usually make various unproven assumptions.
Such an analysis is an important task in order to turn a finite number of perturbative coefficients reliably into very precise estimates for physical quantities and necessary to harvest the predictive power of increasingly high order perturbation series.
Ultimately, we hope that the deep theory of resurgence and trans-series \cite{DunneUensal:CPN-1,Dorigoni:AlienIntroduction}, in combination with the structure of Dyson-Schwinger equations \cite{BellonClavier:DSEBorelSingularities}, will provide superior tools for this task; but so far, its explicit practical lessons seem to restrict to the well-known insight that the analytic continuation of the Borel transform should be tailored to have the expected branch cuts \cite{ChermanKoroteevUnsal:ResurgenceWeakStrong}.

As an application, we resummed the $6$-loop $\eps$-expansions for the critical exponents in $3$ dimensions and found that, in many cases, the resulting reduction of their error estimates renders the renormalization group method again competitive in comparison with recently advanced bootstrap and Monte Carlo techniques.
We expect that the $7$-loop renormalization will provide critical exponents with even higher accuracies and allow for a more stringent analysis of the compatibility of these very different methods.
This is an important task in order to check various assumptions that might be inherent to a particular approach. For example, recent bootstrap results assume that the $O(\n)$ models are realized at a ``kink'' on the boundary of the domain of allowed operator dimensions \cite{3dIsingBootstrapII,EcheverriHarlingSerone:EffectiveBootstrap}.

\begin{acknowledgments}
Both authors are indebted to Kostja Chetyrkin for hospitality at \href{https://www.ttp.kit.edu/}{KIT} in 2014, where this collaboration was started, and for encouraging us to undertake this $6$-loop project.
We furthermore thank Oliver Schnetz for early access to the results of his independent computation, which reassured us on the correctness of our own calculations. The second author is also very grateful for an invitation to \href{https://www.math.uni-erlangen.de/}{FAU Erlangen} and very kind hospitality during this visit.

Dmitrii Batkovich provided valuable checks of many $6$-loop integrals using the IBP/IRR/$\Rstar$-method \cite{BatkovichKompaniets:6loop-Rstar,BatkovichKompaniets:Toolbox}.

Also we thank John Gracey for reminding us of the large $\n$-expansion results \cite{Gracey:LargeNf,BroadhurstGraceyKreimer:PositiveKnots} for the $\beta$ function and for discussions on the resummation of critical exponents. In particular the second author is grateful for hospitality at the \href{https://www.liverpool.ac.uk/mathematical-sciences/seminars/theoretical-physics/}{University of Liverpool}.

David Broadhurst very kindly provided us with a copy of his notes \cite{Broadhurst:WithoutSubtractions}, where he computed the {\MS}-renormalized $5$-loop propagator with extreme economy. The finite parts of these integrals form a subset of the $6$-loop counterterms and provided a valuable check of our results.

Furthermore, we are grateful for Mikhail Nalimov's kind and helpful correspondence regarding the work \cite{KomarovaNalimov:FirstCorrectionO(N),KomarovaNalimov:HigherOrdersO(N)} on the asymptotic expansions in dimensional regularization.
Farrukh Chishtie and Tom Steele very kindly explained to us their work \cite{ChishtieEliasSteele:MassivePhi4} on asymptotic Pad\'{e} approximants and investigated the origin of the huge deviations of the six-loop predictions for $\gamma_{m^2}$ at $\n=5$ (see footnote~\ref{foot:APAP-errors}).

We are also grateful to Andrey Kataev for fruitful discussions.

In this manuscript, we incorporated several suggestions by John, David, Kostja, Mikhail, Oliver and an anonymous referee, who kindly read and commented on earlier drafts.

We thank Simon Liebing for the precise translation of the title of \cite{ChetyrkinGorishnyLarinTkachov:Analytical5loop}.

Also we are grateful to Marc Bellon for discussions on the application of trans-series and resurgence to perturbative quantum field theory and hospitality at \href{https://www.lpthe.jussieu.fr/}{LPTHE Paris}.
Further thanks go to Christopher Beem for an illuminating explanation of the conformal bootstrap program.

The work of the first author was supported by RFBR grant 17-02-00872-a.

The second author's work on this project was supported by the \href{http://www.esi.ac.at/}{ESI Vienna} through the workshop \href{http://www.esi.ac.at/activities/events/2015/the-interrelation-between-mathematical-physics-number-theory-and-non-commutative-geometry}{``The interrelation between mathematical physics, number theory and noncommutative geometry''} and the \href{https://www.mitp.uni-mainz.de/}{Mainz Institute for Theoretical Physics (MITP)} program \href{https://indico.mitp.uni-mainz.de/event/70/}{``Amplitudes: Practical and Theoretical Developments''}. He thanks both institutions for their hospitality during these inspiring meetings.
This project was started while the second author was at the \href{http://www.ihes.fr/}{IH\'{E}S} with support from the ERC grant 257638 via the \href{http://www.cnrs.fr/}{CNRS}.
The first author is grateful to All Souls College and the Mathematical Institute of the University of Oxford for hospitality and support during a visit in 2016.

The calculations of the $Z$-factors and the resummations of the critical exponents were performed on computers of the mathematical institutes of the Humboldt-Universit\"{a}t zu Berlin, the University of Oxford and the Resource Center ``Computer Center of SPbU''.

Figures of Feynman graphs in this article were created with {\JaxoDraw} \cite{BinosiTheussl:JaxoDraw} and {\Axodraw} \cite{Vermaseren:Axodraw}.
\end{acknowledgments}

\appendix

\section{Description of ancillary files}
\label{sec:files}

This article is accompanied by a comprehensive data set in form of human readable text files. We provide these in two formats that are compatible with popular computer algebra software:
{\Maple} \cite{Maple} (files ending with \File{.mpl})
and
{\Mathematica} \cite{Mathematica:11} (files ending on \File{.m}).
Concretely, these include:
\begin{itemize}
	\item $\eps$-expansions of the renormalization group functions and critical exponents,

	\item individual counterterms ($Z$-factor contributions), symmetry and $O(\n)$ factors of all $\leq 6$ loop $\field^4$ graphs and

	\item $\eps$-expansions of the massless propagators that we computed to obtain those.
\end{itemize}
Below we explain in detail the content and format of these files (referring only to the Maple files, since the Mathematica versions are built in complete analogy).

In addition, the attached document \File{resummation.pdf} provides detailed information on our resummations. Namely, for each exponent $f \in \set{\eta,\nu^{-1},\omega}$, dimension $\D\in\set{2,3}$ and the corresponding values of $\n$ considered in tables~\ref{tab:exponents-3d} and \ref{tab:exponents-2d}, it lists the parameters $(b,\lambda,q)$ with least apparent error \eqref{eq:error-estimate} together with plots like in figures~\ref{fig:b-plots-of-lambda} and \ref{fig:lambda-plots-of-b}, showing the nearby dependence on $b$ and $\lambda$, and including the distribution of resummation results as in figure~\ref{fig:resummation-spreads}.

\subsection{RG functions and critical exponents}

Our six loop expansions of the renormalization group functions $\beta,\gamma_{\field}$ and $\gamma_{m^2}$ defined in \eqref{eq:beta-def} and \eqref{eq:anom-dims} are provided in the file \File{expanded.mpl} (in the MS scheme).
It also contains the expansion of the critical coupling $g\crit(\eps)$ from \eqref{eq:gcrit-first-order} and the resulting expansions for the universal critical exponents $\eta$, $\nu^{-1}$ and $\omega$ defined in \eqref{eq:def-omega} and \eqref{eq:def-eta-nu}, plus the critical exponents $\alpha$, $\beta$, $\gamma$ and $\delta$ computed via the scaling relations \eqref{eq:scaling-relations}.

Each expansion is given symbolically with full $\n$-dependence, followed by numeric evaluations for $\n \in \set{0,1,2,3,4}$ like in the tables in sections~\ref{sec:RG-functions} and \ref{sec:resummation}.

\subsection{counterterms of individual graphs}

\begin{figure}
	\centering%
	$
		\NI{\Graph[0.5]{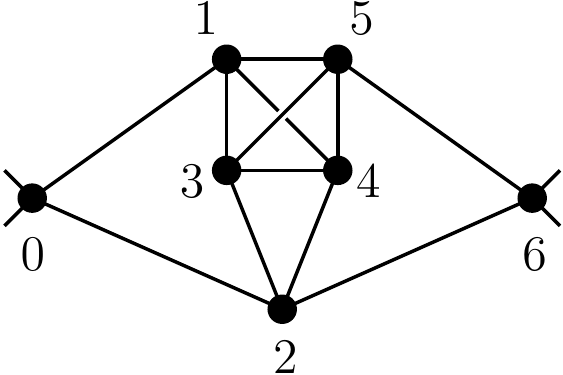}}
		= \texttt{ee12|345|346|45|5|6|ee|}
	$
	\caption{The graph with Nickel index \texttt{ee12|345|346|45|5|6|ee|}, showing the vertex order corresponding to this labelling. This graph gives contributions to $Z_1$ and $Z_4$.}%
	\label{fig:nickel}%
\end{figure}

We computed the $Z$-factors defined in \eqref{eq:Z-factors} via their expansions \eqref{eq:Z-from-R} in counterterms. For each $i\in\set{1,2,4}$, the file \File{Z.mpl} contains a list of graphs contributing to $Z_i$, similar to the table in \cite[Appendix~A]{BatkovichKompanietsChetyrkin:6loop} for $Z_2$.
Each entry is of the form
\begin{equation*}
	\texttt{Z}i\left[ \ell, j \right] \defas
	\left[  \NI{G}, \Sym{G}, \GroupFactor{G}, z_{G}
	\right];
\end{equation*}
and indexed by the loop number $\ell$ and an integer $j$. 
A graph $G$ is specified by its Nickel index $\NI{G}$ as defined in \cite[section~II]{NickelMeironBaker:Compilation24}, see also \cite{BatkovichKirienkoKompanietsNovikov:GraphState}. This is an intuitive notation for an adjacency list with respect to a certain labelling of the vertices, illustrated in figure~\ref{fig:nickel}.
Note that we consider graphs without fixed external labels (``non-leg-fixed'' in the terminology of \cite{Borinsky:feyngen}), i.e.\ the symmetry factor is $\Sym{G} = (E_{G})!/\Aut(G)$ where $E_{G} \in \set{2,4}$ denotes the number of external legs and the automorphisms are allowed to permute them.
Furthermore, the list contains the $O(\n)$ group factor $\GroupFactor{G}$ from \eqref{eq:group-factor-from-vacuum} and the actual counterterm contribution $z_{G}$, which, according to \eqref{eq:Z-from-R}, equals $\partial_{p^2} \Poles \Rbar G$ for $Z_2$ and $\Poles \Rbar G$ for $Z_4$.
The counterterm $Z_1$ is expressed as a linear combination of a subset of graphs contributing to $Z_4$, as explained in section~\ref{sec:computation}.

For example, the entry $\texttt{Z4}[6,458]$ in \File{Z.mpl} corresponds to the graph $G$ depicted in figure~\ref{fig:nickel} with $\NI{G}=\texttt{ee12|345|346|45|5|6|ee|}$.
After minimal subtraction of subdivergences in the $G$-scheme \eqref{eq:G-scheme}, its pole part is
\begin{align}
	\Poles \Rbar \left( \Graph[0.35]{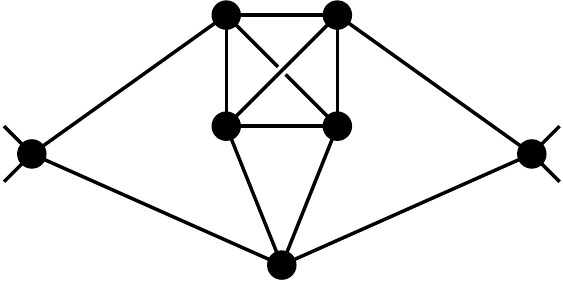} \right)
	&=
	-\frac{\mzv{3}}{3\eps^4}
	+\left(\frac{5}{3}\mzv{3}+\frac{\pi^4}{180} \right)\frac{1}{\eps^3}
	-\left(
		\frac{9}{2}\mzv{3}
		+\frac{7\pi^4}{360}
		-\frac{23}{6}\mzv{5}
	\right)\frac{1}{\eps^2}
\nonumber\\ &\quad
	+\left(
		\frac{9}{2}\mzv{3}
		+\frac{7\pi^4}{360}
		-\frac{161}{30}\mzv{5}
		+\frac{7}{10}\mzv[2]{3}
		-\frac{2\pi^6}{945}
	\right) \frac{1}{\eps}
	.
	\label{eq:ee12_345_346_45_5_6_ee}%
\end{align}
Multiplied with the symmetry and group factors $\Sym{G}=3/2$ and $\GroupFactor{G} = (5\n+22)(3\n^2+22\n+56)/2187$, this gives a contribution to $Z_4$.
The counterterm \eqref{eq:ee12_345_346_45_5_6_ee} also contributes to $Z_1$, but with symmetry factor $-1/2$ and group factor $(\n+2)^2(5\n+22)/243$, as dictated by entry $\texttt{Z1}[6,473]$ in \File{Z.mpl}.
Note that this data can also be looked up in \cite{NickelMeironBaker:Compilation24}, where the graph is called \texttt{603-U7}. Figure~1 ibid.\ also provides drawings of all relevant graphs.

The full $Z$-factor is obtained by summing $ (-g)^{\ell} \Sym{G} \GroupFactor{G} z_{G}$ over all entries of the corresponding list.
The file \File{rg.mpl} demonstrates this computation and furthermore generates the expansions of RG functions and critical exponents from these $Z$-factors (it outputs the contents of \File{expanded.mpl} mentioned earlier).

\subsection{$\eps$-expansions of massless propagators}
\label{sec:appendix-p-integrals}

We obtained the counterterms from massless propagators, which are also called ``$p$-integrals'' \cite{ChetyrkinTkachov:IBP}, via the intermediate use of the BPHZ-like \emph{one-scale scheme} \cite{BrownKreimer:AnglesScales} as explained in detail in \cite{KompanietsPanzer:LL2016}.
Since these $p$-integrals might be valuable for other applications, we include our results for their $\eps$-expansions in the files \File{p\_int\_g$i$.mpl}, where $i\in\set{2,4}$. In the case $i=2$, we list all 1PI propagator graphs of $\field^4$ theory, whereas the $p$-integrals given for $i=4$ arose from nullifying some external momenta and rerouting external legs of some subdivergences, according to \cite{BrownKreimer:AnglesScales,KompanietsPanzer:LL2016}, in order to make those subdivergences single-scale. In particular, this means that the $p$-integrals listed for $i=4$ are usually not Feynman graphs of $\field^4$ theory, due to vertices of valency greater than four.

The $\eps$-expansions are given for external momentum squared $p^2=1$ in the $G$-scheme \cite{ChetyrkinKataevTkachov:Gegenbauer}: each integration over a loop momentum $\vec{k}$ carries the measure
\begin{equation}
	\frac{\Gamma^2(1-\eps) \Gamma(1+\eps)}{\Gamma(2-2\eps)}
	\int \frac{\dd[4-2\eps] \vec{k}}{\pi^{2-\eps}}
	,
	\label{eq:G-scheme}%
\end{equation}
which in particular normalizes the bubble $\NI{\Graph[0.3]{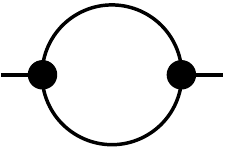}} = \texttt{e11|e|}$, which is the first entry $\texttt{pInt}[1,1]$ in \File{p\_int\_4.mpl}, to be exactly $1/\eps$.
For example, the entry $\texttt{pInt}[6,163]$ with Nickel index $\texttt{e123|e24|35|66|56|6||}$ gives the $\eps$-expansion
\begin{equation}
	\left[ \Graph[0.28]{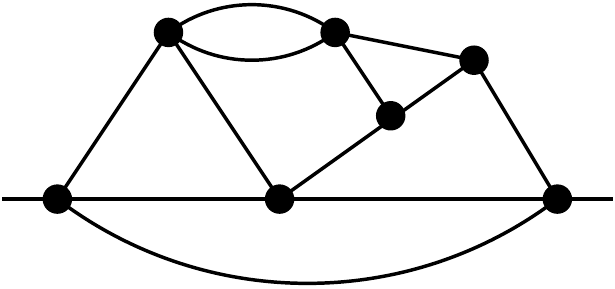} \right]_{p^2=1}
	=
	\frac{147\mzv{7}}{16\eps^2}
	-\left(
		\frac{147}{16} \mzv{7}
		+\frac{27}{2}\mzv{3}\mzv{5}
		+\frac{27}{10}\mzv{3,5}
		-\frac{2063\pi^8}{504000}
	\right)\frac{1}{\eps}
	+\bigo{\eps^0}.
	\label{eq:e123_e24_35_66_56_6}%
\end{equation}

\section{Estimates for primitive diagrams up to 11 loops}
\label{sec:Hepp-bound}

In the discussion of the asymptotic behaviour in section~\ref{sec:asymptotics}, we included the contributions of primitive (subdivergence free) graphs to the $\beta$ function with up to $11$ loops in figure~\ref{fig:beta-asymptotics}.
These graphs had been enumerated in \cite{Schnetz:Census}, but exact results are currently complete only up to $7$ loops \cite{PanzerSchnetz:Phi4Coaction}. The Feynman integral of such a primitive $4$-point graph $G$ with $\ell$ loops (and $\ell+1$ vertices) has a simple pole $\Period{G}/(\ell\eps)$ and thus contributes $-(-g)^{\ell} \Period{G}/\ell$ to $Z_4$ and $Z_g$.
Taking the symmetry factors into account, the resulting contributions $\beta^{\prim}(g) = \sum_k \beta^{\prim}_k (-g)^k$ to the beta function defined in \eqref{eq:beta-def} are
\begin{equation}
	\beta^{\prim}_{\ell+1}
	= 2\times \sum_{\substack{\text{primitive $G$,}\\\text{$\ell$ loops, $4$ legs}}}
	\frac{\Period{G}}{\abs{\Aut(G)}}
	.
	\label{eq:primitive-beta}
\end{equation}
The residue $\Period{G}$ is known as the \emph{Feynman period} \cite{BEK} and can be written in Schwinger parameters $\SP_e$ (one for each edge $e\in E(G)$ of the graph) as
\begin{equation}
	\Period{G}
	= \int_0^{\infty} \!\!\dd \SP_2 \ \cdots \int_0^{\infty}\!\! \dd \SP_{E(G)}\ \frac{1}{\restrict{\psipol_G^2(\SP)}{\SP_1=1}}
	.
	\label{eq:period}%
\end{equation}
Here, the \emph{Symanzik polynomial} $\psipol_G$ is given by a sum over spanning trees $T$ of $G$,
\begin{equation}
	\psipol_G
	= \sum_{\substack{\text{spanning}\\\text{tree $T$}}} \ \prod_{e \notin T} \SP_e
	.
	\label{eq:psi}%
\end{equation}
Beyond $7$ loops, not all Feynman periods are currently known \cite{PanzerSchnetz:Phi4Coaction}.
Standard numerical techniques for the evaluation of \eqref{eq:period} are based on sector decomposition and Monte Carlo integration \cite{BinothHeinrich:SectorDecomposition}. Unfortunately, these methods are not applicable in practice to graphs as complicated as the $\field^4$ graphs with $11$ loops that we are facing here.
However, we find that a decent numerical estimate for the integrals \eqref{eq:period} can be obtained from a rather simple graph invariant, constructed by approximating the Symanzik polynomial \eqref{eq:psi} with its dominant (maximal) monomial.\footnote{%
	The integration domain can be subdivided into \emph{Hepp sectors} $\setexp{x}{x_{\sigma(1)}<\cdots<x_{\sigma(E(G))}}$, which are indexed by a permutation $\sigma$ of the edges $E(G)$. Inside each Hepp sector, a particular monomial of $\psipol_G$ dominates all other monomials.
}
\begin{definition}
	\label{def:Hepp}%
	The \emph{Hepp bound} of a primitive $\phi^4$ graph $G$ is
	\begin{equation}
		\Hepp{G}
		\defas \int_0^{\infty}\!\! \dd \SP_2\ \cdots \int_0^{\infty}\!\! \dd \SP_{E(G)}\ 
		\frac{1}{\restrict{\left( \max_T \prod_{e \notin T} \SP_e \right)^2}{\SP_1=1}}
		\in \Q
		.
		\label{eq:def-Hepp}%
	\end{equation}
\end{definition}
Note that $\Hepp{G} \geq \Period{G}$ is indeed an upper bound. It is much easier to compute than the actual period~\eqref{eq:period}; in particular, it is just a rational number and the calculation of the Hepp bounds for all primitive graphs with up to $11$ loops is possible without much difficulty.
The Hepp bound has many more interesting properties and will be explored in detail in a paper by the second author~\cite{Panzer:HeppBound}.

\begin{figure}
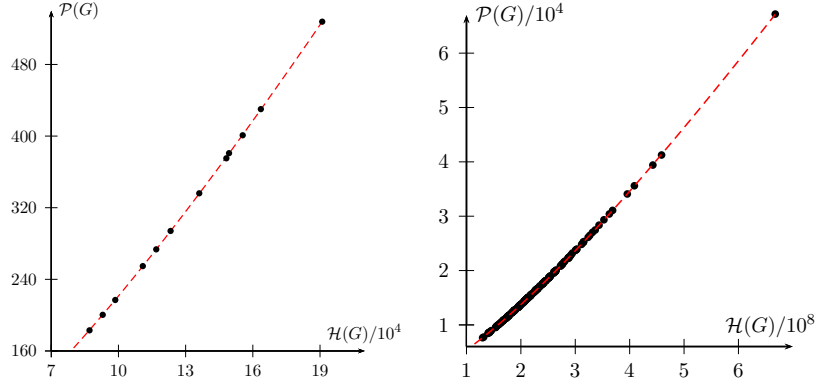

	\centering
	\includegraphics[height=5cm]{{{plots/hepp.7loops}}}
	\quad
	\includegraphics[height=5cm]{{{plots/hepp.11loops}}}
	\caption{The known Feynman periods $\Period{G}$ at $7$ loops (left) and $11$ loops (right) as a function of the Hepp bound $\Hepp{G}$ from definition~\ref{def:Hepp}. The plots also show the interpolating functions of the form \eqref{eq:Hepp-fit}.}%
	\label{fig:Hepp-bounds}%
\end{figure}

What is relevant for our purposes here is that, surprisingly, these easily obtainable numbers correlate strongly with the complicated Feynman period. For example, figure~\ref{fig:Hepp-bounds} shows the period $\Period{G}$ as a function of the Hepp bound $\Hepp{G}$ at $7$ loops, where all periods are known exactly due to \cite{PanzerSchnetz:Phi4Coaction}.

This relationship between $\Period{G}$ and $\Hepp{G}$ is well approximated by a power law, and even better when we allow for one further parameter as in
\begin{equation}
	\Period{G} \approx f_{\ell}\left(\Hepp{G}\right) 
	\quad\text{where}\quad
	f_{\ell}(h) \defas a_{\ell} h^{b_{\ell}} \left( 1 - h c_{\ell} \right).
	\label{eq:Hepp-fit}%
\end{equation}
We fit the parameters $a_{\ell}$, $b_{\ell}$ and $c_{\ell}$ to the known periods at loop order $\ell$. The resulting coefficients are given in table~\ref{tab:primitive-estimates} and the plots of $f_{\ell}$ are shown in figure~\ref{fig:Hepp-bounds} for $\ell=7$ and $\ell=11$ loops.
In all cases, the approximation by the very simple fit curves \eqref{eq:Hepp-fit} reproduces the known periods within $2\%$ accuracy.\footnote{%
	This is a bound on the error for an individual period $\Period{G}$ as approximated by $f_{\ell}\left( \Hepp{G} \right)$. On average, the error is much smaller.
}
We note though that, with growing loop number, only very few periods are known exactly, as current integration techniques only apply to graphs with certain special combinatorial structures \cite{PanzerSchnetz:Phi4Coaction}. Our fits are thus biased---however, for the purpose of our discussion here, we expect this systematic error to be negligible.

To estimate the primitive contributions $\beta^{\prim}_{\ell+1}$ to $\beta^{\MS}_{\ell+1}$ at loop order $\ell$, we substitute $f_{\ell}\left( \Hepp{G} \right)$ for $\Period{G}$ in \eqref{eq:primitive-beta}. The calculation can be economized due to the fact that primitive graphs $G$ with the same completion $F$ (the completion is the 4-regular graph obtained by adding a vertex $v$ and connecting it to the external legs, such that $G=F\setminus v$) have equal periods \cite{Schnetz:Census} and Hepp bounds. Therefore, $\Hepp{F} \defas \Hepp{F\setminus v}$ is independent of the choice of the vertex $v$.
Table~\ref{tab:primitive-estimates} summarizes the number of (isomorphism classes of) such completions, which were given also in \cite[Table~1]{Schnetz:Census}.
We used {\nauty}~\cite{McKayPiperno:II} to generate these graphs and to count their automorphisms.

\begin{table}
\caption{%
	The upper part shows the counts of completed primitive $\field^4$ graphs and the estimates for the primitive contributions $\beta_{\ell+1}^{\prim}$ to the beta function of the $\field^4$ model ($\n=1$) at $\ell$-loop order obtained with the Hepp bounds \eqref{eq:def-Hepp}.
	The lower part of the table shows the fitting parameters used in the approximation \eqref{eq:Hepp-fit}.
}%
	\label{tab:primitive-estimates}%
	\centering
	\begin{tabular}{rrrrrrr}
	\toprule%
		loop order $\ell$ & 6 &  7 &  8 &   9 &   10 &   11 \\ 
	\midrule%
		completions       & 5 & 14 & 49 & 227 & 1354 & 9722 \\ 
		$\beta^{\prim}_{\ell+1}$ estimate 
				  & $2.41 \!\cdot\! 10^4$  & $3.71\!\cdot\! 10^5$   & $6.06 \!\cdot\! 10^6$   & $1.05 \!\cdot\! 10^8$    &  $1.89 \!\cdot\! 10^9$    &  $3.57 \!\cdot\! 10^{10} $\\
		$\beta^{\prim}_{\ell+1}/\overline{\beta}^{\MS}_{\ell+1}$
				  & 21.8\% & 26.2\% & 31.6\% & 37.6\% & 44.3\% & 51.5\% \\		  
	\midrule%
	$a_{\ell}$        & $9.78/10^{5}$ & $2.44/10^{5}$ & $5.22/10^{6}$ & $1.16/10^{6}$ & $2.34 / 10^{7}$ & $5.11 /10^{8}$ \\
		$b_{\ell}$        &1.419 & 1.395 & 1.389 & 1.382 & 1.382 & 1.378 \\
		$c_{\ell}$        & $2.93 /10^{6}$ & $3.48/ 10^{7}$ & $4.98 / 10^{8}$ & $6.87 / 10^{9}$ & $1.04 / 10^{9}$ & $1.46 / 10^{10}$ \\
	\bottomrule%
	\end{tabular}
\end{table}

Using the orbit-stabilizer theorem, $\abs{\Aut(F\setminus v)} = \abs{ \mathrm{Stab}_{\Aut(F)}(v)} = \abs{\Aut(F)} / \abs{\Aut(F) \cdot v}$, we can express the primitive contributions \eqref{eq:primitive-beta} as a sum over the completed graphs as\footnote{%
	The factor $4!$ accounts for the different labellings of the external legs of a decompletion, and the $\ell+2$ vertices $V(F)= \mathbin{\dot{\bigcup}}_{\text{orbits}} \Aut(F) \cdot v$ are partioned into the orbits corresponding to inequivalent uncompletions $G=F\setminus v$.%
}
\begin{equation}
	\beta^{\prim}_{\ell+1} (\n)
	\approx 
	2\cdot
	4!\cdot(\ell+2)
	\times
	\sum_{\substack{\text{primitive $F$,}\\\text{$\ell$ loops, $4$-reg.}}}
	\frac{f_{\ell}\left( \Hepp{F} \right)}{\abs{\Aut{(F)}}}
	\cdot \frac{3 \GroupFactor{F}}{\n(\n+2)}
	,
	\label{eq:sum-over-primitive-completions}%
\end{equation}
where we also made the group factor $\GroupFactor{F}$ from \eqref{eq:group-factor} explicit, using \eqref{eq:group-factor-from-vacuum}.
In table~\ref{tab:primitive-estimates} we show the results for $\n=1$ and see, for example, that at $11$ loops, the primitive contributions amount to $51.5\%$ of the asymptotic prediction $\overline{\beta}^{\MS}_{12}$ from \eqref{eq:beta-asymptotics}. Based on a comparison in the $7$ loop case, where we know $\beta^{\prim}_8$ exactly, we expect our estimates for $\beta^{\prim}_{\ell+1}$ to be accurate within 1\textperthousand.

\section*{Bibliography}

\bibliographystyle{JHEPsortdoi}
\bibliography{refs}

\end{document}